\documentclass[twocolumn,english,superscriptaddress,notitlepage]{revtex4-1}
\usepackage[T1]{fontenc}
\usepackage[latin9]{inputenc}
\usepackage{graphicx}
\usepackage{amsmath}
\usepackage[usenames, dvipsnames]{color}

\makeatletter
\usepackage{babel}

\makeatother

\begin{document}

\title{Crystal Field Levels and Magnetic Anisotropy in the Kagome Compounds $\rm{Nd_3Sb_3Mg_2O_{14}}$,  $\rm{Nd_3Sb_3Zn_2O_{14}}$, and  $\rm{Pr_3Sb_3Mg_2O_{14}}$}

\author{A. Scheie}
\address{Institute for Quantum Matter and Department of Physics and Astronomy, Johns Hopkins University, Baltimore, MD 21218}

\author{M. Sanders}
\address{Department of Chemistry, Princeton University, Princeton, NJ 08544 }

\author{J. Krizan}
\address{Department of Chemistry, Princeton University, Princeton, NJ 08544 }

\author{A. D. Christianson}
\address{Neutron Scattering Division, Oak Ridge National Laboratory, Oak Ridge, Tennessee 37831, USA}
\address{Materials Science \& Technology Division, Oak Ridge National Laboratory, Oak Ridge, TN 37831, USA}

\author{V. O. Garlea}
\address{Neutron Scattering Division, Oak Ridge National Laboratory, Oak Ridge, Tennessee 37831, USA}

\author{R.J. Cava}
\address{Department of Chemistry, Princeton University, Princeton, NJ 08544 }

\author{C. Broholm}
\address{Institute for Quantum Matter and Department of Physics and Astronomy, Johns Hopkins University, Baltimore, MD 21218}
\address{NIST Center for Neutron Research, National Institute of Standards and Technology, Gaithersburg, MD 20899}
\address{Department of Materials Science and Engineering, Johns Hopkins University, Baltimore, MD 21218}

\date{\today}

\begin{abstract}
We report the crystal field levels of several newly-discovered rare-earth kagome compounds: $\rm{Nd_3Sb_3Mg_2O_{14}}$,  $\rm{Nd_3Sb_3Zn_2O_{14}}$, and  $\rm{Pr_3Sb_3Mg_2O_{14}}$. We determine the CEF Hamiltonian by fitting to neutron scattering data using a point-charge Hamiltonian as an intermediate fitting step. The fitted Hamiltonians accurately reproduce bulk susceptibility measurements, and the results indicate easy-axis ground state doublets for $\rm{Nd_3Sb_3Mg_2O_{14}}$ and $\rm{Nd_3Sb_3Zn_2O_{14}}$, and a singlet ground state for  $\rm{Pr_3Sb_3Mg_2O_{14}}$. These results provide the groundwork for future investigations of these compounds and a template for CEF analysis of other low-symmetry materials. 

\end{abstract}

\maketitle

\section{Introduction}

The kagome lattice of corner-sharing triangles is the basis for multiple distinct  forms of frustrated magnetism with unique physical properties. Magnetic kagome lattices are believed to host spin-liquid phases \cite{Essafi2016, Singh1992, Yan2011, Gong2015}, non-trivial transport properties \cite{Hirschenberger2015}, and topologically protected phases \cite{Ohgushi2000}. Experimental realizations of these models present important opportunities to explore new states of matter.

Recently, a new family of kagome compounds with magnetic rare earth ions $\rm{RE_3Sb_3A_2O_{14}}$ (RE = rare earth, A = Mg, Zn) was discovered \cite{Dun2016, SandersREMg, SandersREZn}. 
Basic materials characterization has been carried out on the entire family \cite{Dun2016, Dun2017, SandersREMg, SandersREZn}, and neutron diffraction has revealed the low temperature magnetic structure of $\rm{Nd_3Sb_3Mg_2O_{14}}$ \cite{MyPaper} and $\rm{Dy_3Sb_3Mg_2O_{14}}$ \cite{Paddison2016}. $\mu$SR data for $\rm{Tb_3Sb_3Zn_2O_{14}}$ were interpreted as indicative of a spin-liquid ground state \cite{Ding2018}. 

\begin{figure} 
	\centering\includegraphics[scale=0.09]{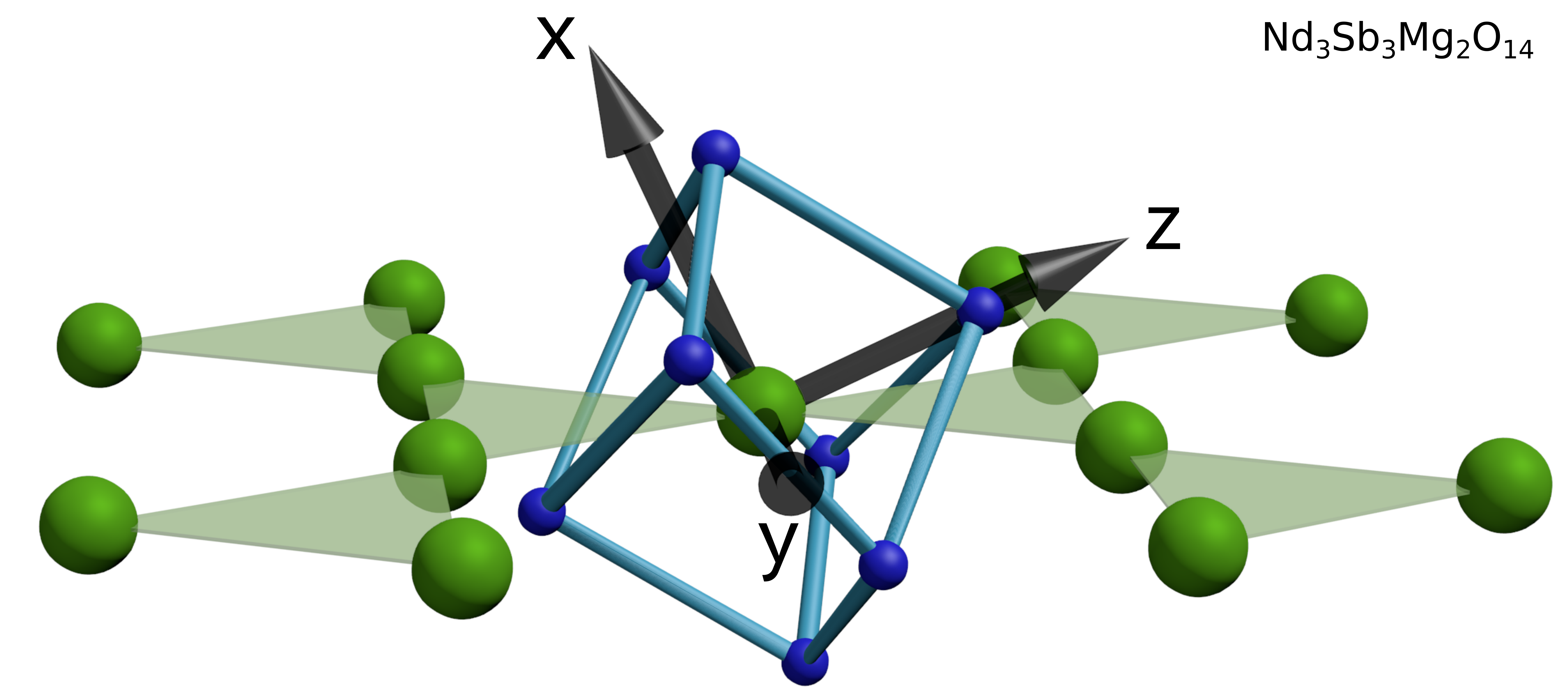}
	
	\caption{Distorted scalenohedron ligand environment of Nd in $\rm{Nd_3Sb_3Mg_2O_{14}}$. The axes at the center show the local axes used to model the crystal fields. $\rm{Nd_3Sb_3Zn_2O_{14}}$ and $\rm{Pr_3Sb_3Mg_2O_{14}}$ have the same symmetry, but with slightly different oxygen locations.
	}
	\label{flo:LigandEnvt}
\end{figure}

The rare earth ions in these materials are strongly influenced by the electrostatic environment they occupy. It determines to what extent and how the $2J+1$ fold spin-orbital degeneracy of the rare earth ion is lifted \cite{AbragamBleaney}.  Clearly this has major impacts on the nature of the potentially frustrated magnetism. Fortunately the crystal electric field (CEF) level scheme can be accurately determined using inelastic neutron scattering and it is to this task that we have devoted ourselves in this paper. Specifically, we report the crystal field Hamiltonians of $\rm{Nd_3Sb_3Mg_2O_{14}}$,  $\rm{Nd_3Sb_3Zn_2O_{14}}$, and  $\rm{Pr_3Sb_3Mg_2O_{14}}$ deduced from crystal field excitations observed with neutron scattering.
The complexity of the CEF Hamiltonian is determined by the point group symmetry of the ion: high symmetry means few CEF parameters, low symmetry means many CEF parameters. The ligand environment for $\rm{RE_3Sb_3A_2O_{14}}$ (RE = rare earth, A = Mg, Zn) has a very low symmetry oxygen environment of $2/m$ symmetry (see Fig. \ref{flo:LigandEnvt}), leading to 13 allowed CEF parameters in the Hamiltonian. Such a model is very difficult to uniquely establish, but---by using a point-charge approximation to obtain a first approximation---reliable fits to neutron scattering data are possible.
The techniques outlined here provide a template for analyzing the rest of this family of Kagome compounds and indeed they should be useful for analyzing the crystal field level scheme when as here the symmetries involved are low. 

\section{Experimental Methods}

We performed neutron scattering experiments on 5g of $\rm{Nd_3Sb_3Mg_2O_{14}}$, 5g $\rm{Nd_3Sb_3Zn_2O_{14}}$, and 5g  $\rm{Pr_3Sb_3Mg_2O_{14}}$ (all loose powders) on the ARCS spectrometer at the SNS at ORNL. For every compound, we collected data at incident energies $E_i=150\>$meV, $E_i=80\>$meV, and $E_i=40\>$meV; at temperatures $T=6\>$K, $T=100\>$K, and $T=200\>$K for every $E_i$ (a total of nine data sets), measuring for two hours at each setting. We also acquired data for a 5g nonmagnetic analogue $\rm{La_3Sb_3Mg_2O_{14}}$ to serve as a background. This allows us to subtract the phonon contribution from the data for the magnetic compounds (see supplemental materials for details about the background subtraction). 

\begin{figure*} 
	\centering\includegraphics[scale=0.72]{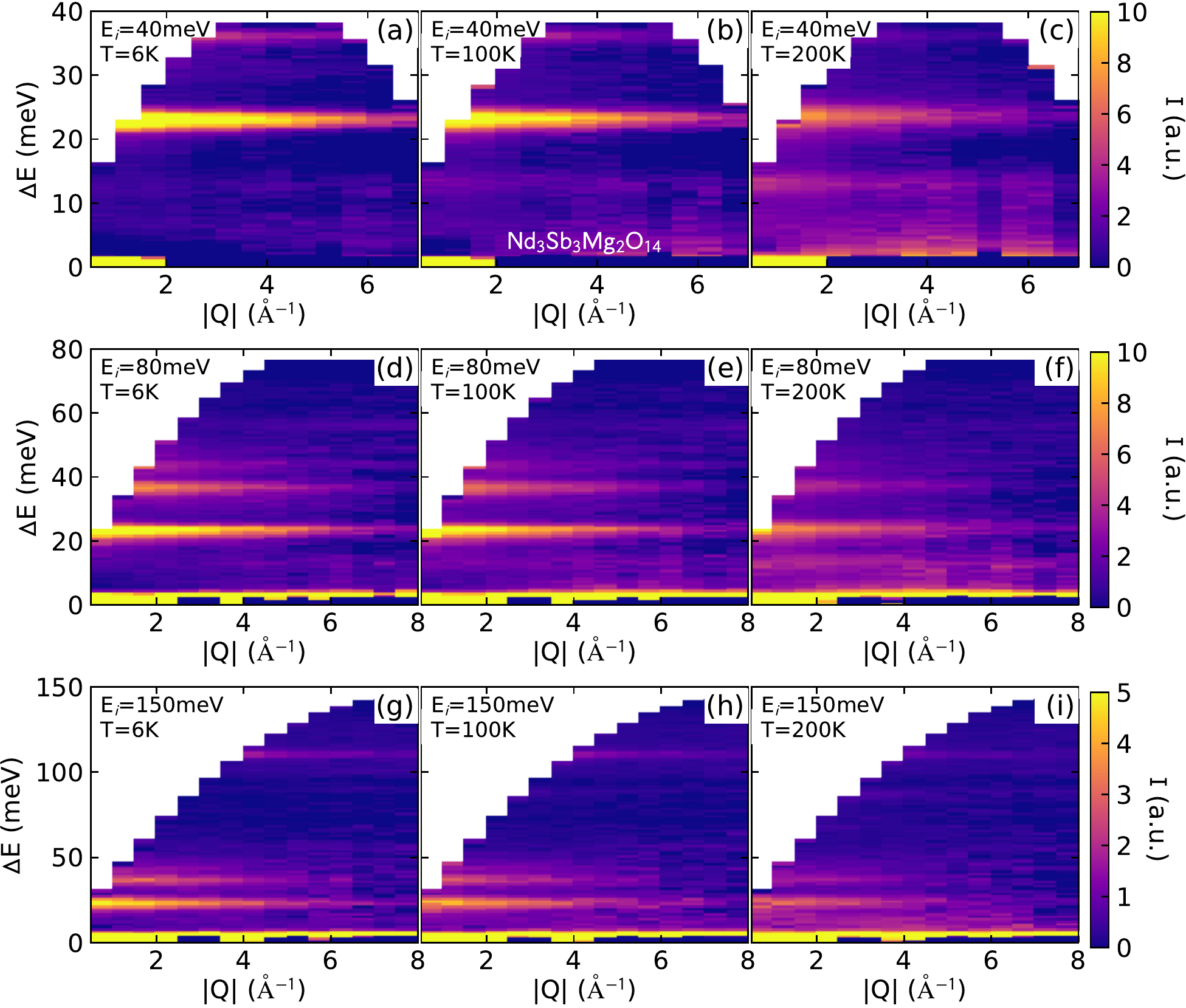}
	
	\caption{Inelastic neutron scattering from $\rm{Nd_3Sb_3Mg_2O_{14}}$, taken at $E_i=$150 meV, 80 meV, and 40 meV and $T=$6 K, 100 K, and 200 K. Scattering from nonmagnetic $\rm{La_3Sb_3Mg_2O_{14}}$ was scaled and subtracted to eliminate phonon scattering. The CEF excitations are clearly visible and become broadened as temperature increases.
	}
	\label{flo:NdMg_NeutronData}
\end{figure*}

The full background-subtracted data set for $\rm{Nd_3Sb_3Mg_2O_{14}}$ is shown in Fig. \ref{flo:NdMg_NeutronData}. The crystal field excitations are clearly visible because the corresponding intensity decreases  with $Q$ as a result of the electronic form factor. As Nd$^{3+}$ is a $J=9/2$ Kramers ion, we expect to see $10/2=5$ CEF levels, and thus four CEF transition energies from the ground state. This is indeed what we observe in the neutron data: in the 6 K data, four transitions are visible at 23 meV, 36 meV, 43 meV, and 111 meV.   At higher temperatures, the existing peaks broaden in $\Delta E$ due to shorter excited-state lifetimes, and additional weak peaks appear corresponding to transitions between thermally populated excited levels.

\begin{figure*} 
	\centering\includegraphics[scale=0.56]{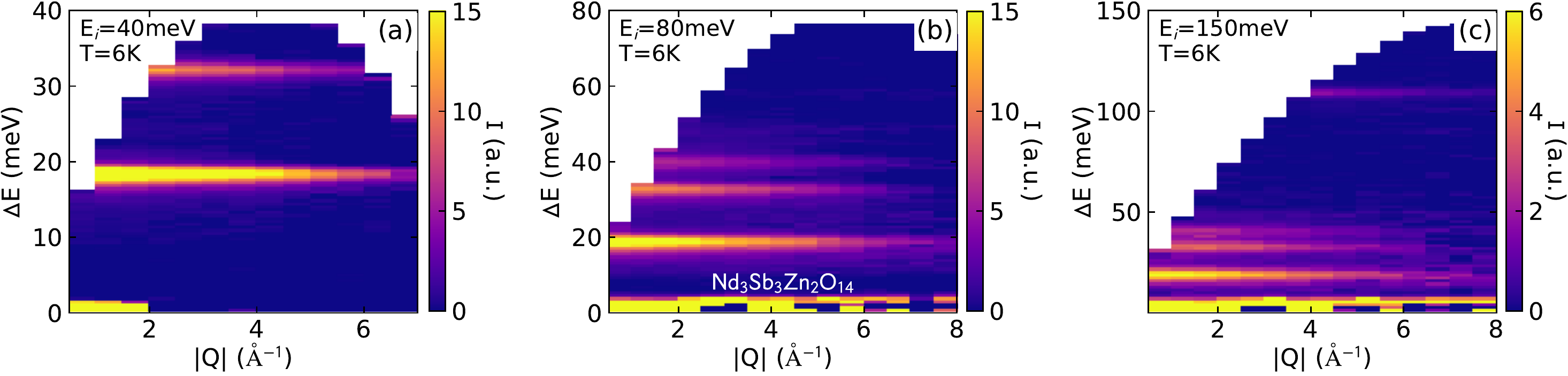}
	
	\caption{Inelastic neutron scattering from $\rm{Nd_3Sb_3Zn_2O_{14}}$, taken at $E_i=$150 meV, 80 meV, and 40 meV and $T=$6 K (100 K and 200 K data are not shown). Scaled $\rm{La_3Sb_3Mg_2O_{14}}$ scattering was subtracted. Note the similarity of the patterns to Fig. \ref{flo:NdMg_NeutronData}.}
	\label{flo:NdZn_NeutronData}
\end{figure*}

An abbreviated (6 K only) data set for $\rm{Nd_3Sb_3Zn_2O_{14}}$ is shown in Fig. \ref{flo:NdZn_NeutronData}. These data are nearly identical to the  $\rm{Nd_3Sb_3Mg_2O_{14}}$ data in Fig. \ref{flo:NdMg_NeutronData}, with four transitions from a $J=9/2$ Kramers ion, but with transition energies at 18 meV, 32 meV, 40 meV, and 109 meV. Such differences indicate slight modifications in the ligand environment experienced by the rare earth ion.

\begin{figure*} 
	\centering\includegraphics[scale=0.56]{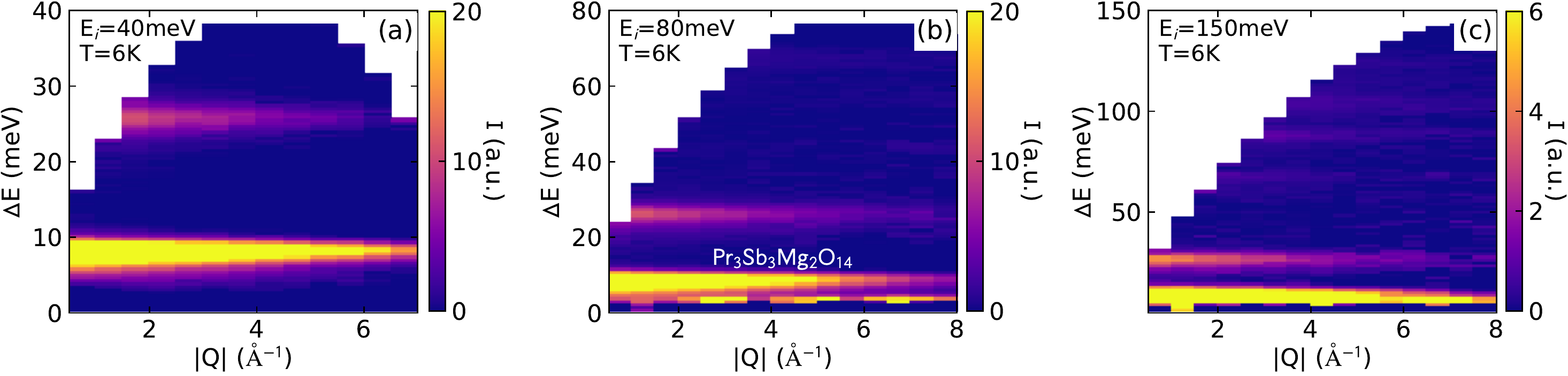}
	
	\caption{Inelastic neutron scattering from $\rm{Pr_3Sb_3Mg_2O_{14}}$, taken at $E_i=$150 meV, 80 meV, and 40 meV and $T=$6 K (100 K and 200 K data are not shown). Scaled $\rm{La_3Sb_3Mg_2O_{14}}$ scattering was subtracted. There is a very strong transition at 7.5 meV, and much weaker transitions at higher energies.
	}
	\label{flo:PrMg_NeutronData}
\end{figure*}

An abbreviated (6 K only) data set for $\rm{Pr_3Sb_3Mg_2O_{14}}$ is shown in Fig. \ref{flo:PrMg_NeutronData}. Pr$^{3+}$ is a non-Kramers ion with $J=4$, which means that singlet states are possible when the point group symmetry is sufficiently low so the number of transitions observed is greater. Five transitions are clearly  distinguishable at 6 K, with more being too weak to distinguish in the figure.

\section{Computational Methods}

Using the inelastic neutron scattering data, we were able to infer a crystal field model for each of the compounds that can account for their anisotropic magnetic properties for temperatures above the inter-site interaction scale (1 K). The fits were carried out using the PyCrystalField software package \cite{PyCrystalField}. The analysis is based on the following CEF Hamiltonian
\begin{equation}
\mathcal{H}_{CEF} =\sum_{n,m} B_{n}^{m}O_{n}^{m}.
\end{equation}
Here $O_{n}^{m}$ are the Stevens Operators \cite{Stevens1952,Hutchings1964} and $B_{n}^{m}$ are multiplicative factors called CEF parameters that parametrize the effects of the ligand environment on the rare earth ion. This formalism is convenient when the ligand environment has high symmetry, leaving only a handful of CEF parameters to be fit \cite{Hutchings1964}.
Unfortunately, a direct fit to the data for $\rm{Nd_3Sb_3Mg_2O_{14}}$ is not feasible: fitting 13 parameters to eight observables (four transition energies and four neutron intensities). To get around this, we begin with a constrained fit based on an electrostatic point-charge model of the ligand environment. Specifically, the point charge $\mathcal{H}_{CEF}$ is based on a Taylor expansion of the electrostatic field at the rare earth site generated by the coordinating atoms treated as point charges \cite{Hutchings1964,Mesot1998}.

Following the method outlined by Hutchings \cite{Hutchings1964}, the CEF parameters $B_n^m$ are given by
\begin{equation}
B_{n}^{m}=-\gamma_{nm}q~C_{nm}\left\langle r^{n}\right\rangle \theta_{n}.
\label{eq:CEFparams}
\end{equation}
Here $\gamma_{nm}$ is a term calculated from the ligand environment expressed in terms of tesseral harmonics, $q$ is the charge of the central ion (in units of $|e|$), $C_{nm}$ are normalization factors of the tesseral harmonics \cite{Hutchings1964}, $\left\langle r^{n}\right\rangle$ is the expectation value of the radial wavefunction for the rare earth ion \cite{EDVARDSSON1998}, and $\theta_{n}$ are multiplicative factors from expressing the electrostatic potential in terms of Stevens Operators in the $J$ basis \cite{Stevens1952}. 

The neutron cross section for a single CEF transition in a powder sample is
\begin{multline}
\frac{d^2\sigma}{d\Omega d\omega} = N (\gamma r_0)^2 \frac{k'}{k}f^2(\mathbf{Q}) e^{-2W(\mathbf{Q})} \\
p_n|\langle \Gamma_m|\hat J_{\perp}|\Gamma_n \rangle|^2 \delta(\hbar \omega + E_n - E_m)
\label{eq:NeutronCrossSec}
\end{multline}
\cite{Furrer2009neutron}, where $N$ is the number of ions, $\gamma = 1.832 \times 10^{8} {\rm s^{-1} T^{-1}}$ is the gyromagnetic ratio of the neutron, $r_0=2.818 \times 10^{-15} {\rm m}$ is the classical electron radius, $k$ and $k'$ are the incoming and outgoing neutron wavevectors, $f(\mathbf{Q})$ is the form factor, $e^{-2W(\mathbf{Q})}$ is the Debye Waller factor, $p_n = e^{-\beta E_n} / \sum_i e^{-\beta E_i}$ is the Boltzmann weight, and $|\langle \Gamma_m|\hat J_{\perp}|\Gamma_n \rangle|^2  = 
\frac{2}{3} \sum_{\alpha} |\langle \Gamma_m|\hat J_{\alpha}|\Gamma_n \rangle|^2 $ is computed from the inner product of total angular momentum $J_{\alpha}$ with the CEF eigenstates $|\Gamma_n \rangle$. Using this equation, one can calculate the neutron spectrum of a given CEF Hamiltonian at a given temperature.
In reality, the delta function $\delta(\hbar \omega + E_n - E_m)$ is replaced with a finite width peak due to the limited energy resolution of the instrument, dispersion, and/or the finite lifetime of the excitation. The resolution was approximated with a Gaussian profile, while finite lifetimes give Lorentzian profiles. We approximated the convolution of these with a Voigt profile for computational efficiency. The energy transfer depedendent resolution width was calculated as described in ref. \cite{ARCS_resolution} with sample width $dL_3$ defined so the calculated Full Width at Half Maximum (FWHM) of the elastic line matched the measured FWHM. The finite lifetime Lorentzian width was a single temperature dependent fitting parameter shared by all transitions.

\begin{figure}
	\centering\includegraphics[scale=0.45]{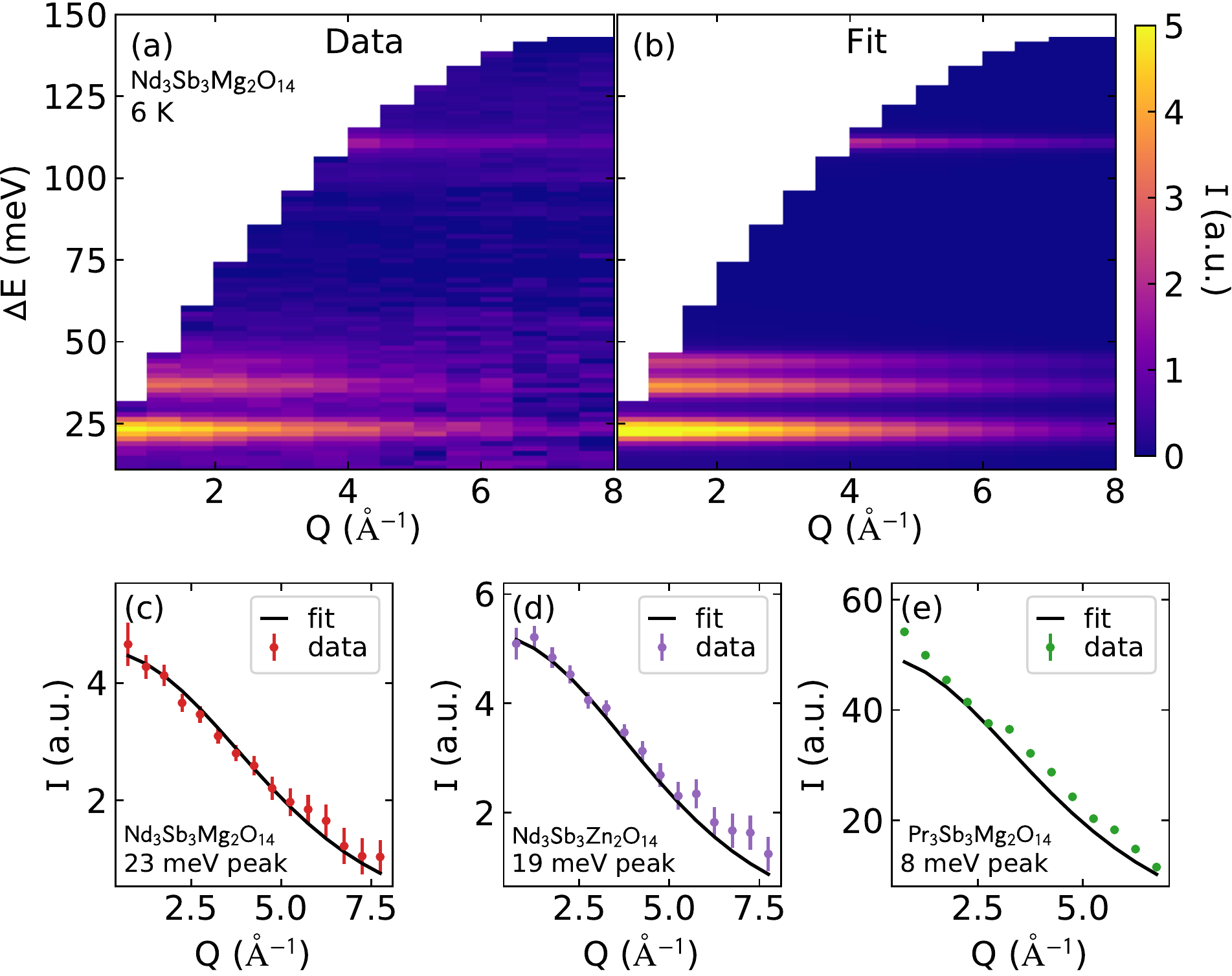}
	
	\caption{Example of a 2D fit to neutron scattering data. (a) and (b) show the data and final fit results of the 2D data set for $\rm{Nd_3Sb_3Mg_2O_{14}}$ at 6 K. Panels (c) - (e) compare the calculated and observed Q dependence for the integrated intensity of the lowest energy excitation peaks at 6 K for the three compounds.
	}
	\label{flo:Sample2D_fit}
\end{figure}

Multiple constant-$Q$ spectra were fitted simultaneously computing the $Q$-dependent scattering using the calculated form factor and a temperature dependent Debye-Waller factor approximated with an overall thermal parameter $u$ \cite{Squires} (see supplemental materials for details). We also fit simultaneously to data at all energy transfers and temperatures, for a total of nine $Q$ and $\Delta E$ dependent data sets being fit simultaneously for each compound. For $E_i=150$ meV and $E_i=80$ meV, we fit data up to 8 \AA$^{-1}$, and for $E_i=40$ meV we fit up to 7 \AA$^{-1}$ (at which points the magnetic intensity was indistinguishable from background noise).

Using the point charge formalism described above, we fit the CEF Hamiltonians in three steps. The first step was calculating the CEF parameters for each compound using the ligand positions refined in refs. \cite{MyPaper,SandersREMg,SandersREZn}. (We refer to this as the "Calculated PC" model.)
As a second step, we refined the effective charges of the symmetry-independent ligand sites by fitting the calculated neutron spectrum to the data. (We refer to this as the "PC Fit" model.) In $\rm{RE_3Sb_3A_2O_{14}}$, there are eight ligands surrounding RE but only three symmetry-independent ligand sites. So we fit the effective charges (contained in $\gamma_{nm}$) of each symmetry-independent atom, thus fitting the relative weights of each symmetry-related group of ligands, starting with effective charges of $(-2e, -2e, -2e)$ for O$^{2-}$ ions. By fitting effective charges, we have three fitted parameters and eight observables. 
In fitting the effective point charge model, we added a term to the global $\chi^2$ measuring the mean square deviation of the calculated transitions from the observed transitions (which were taken from Gaussian fits to the spectra) of the form $\chi^2_{\Delta E} = \sum_i (E^{obs}_i-E^{calc}_i)^2$. This was found to improve convergence.

As a third and final step, we used the crystal field parameters $B_n^m$ obtained from the best effective charge fit as starting parameters for a fit to neutron data varying all $B_n^m$. (We refer to this as the "Final Fit" model.) We included a weakly weighted $\chi^2_{\Delta E}$ term in the final fit to keep the fit from wandering astray. In doing so, we assume that the point charge fit approaches the global minimum in $\chi^2$ and that the final fit is merely an adjustment to the best-fit point charge model.

To cross-check our results, we computed the magnetic susceptibility from $\mathcal{H}_{CEF}$ numerically. Susceptibility is defined as $\chi_{\alpha, \beta} = \frac{\partial M_{\alpha}}{\partial H_{\beta}}$, and $M_{\alpha} = g_J \langle J_{\alpha} \rangle$, where $\langle J_{\alpha} \rangle = \sum_i e^{\frac{-E_i}{k_B T}}\langle i \rvert J_{\alpha} \lvert i \rangle ~/ Z~$ and $|i\rangle$ are the eigenstates of the effective Hamiltonian ${\cal H} = {\cal H}_{CEF} + \mu_B g_J \mu_0{\bf H}\cdot {\bf J} $, where $\mu_0{\bf H}$ is magnetic field. Computing $M_{\alpha}$ at various fields and taking a numerical derivative with respect to field yields the magnetic susceptibility. Figure 9 provides a comparison of the calculated powder-average susceptibility compared with experimental data. The calculated anistropic low-temperature magnetization is in Fig. \ref{flo:Anisotropy}.

\section{Results}

\subsection{$\rm{Nd_3Sb_3Mg_2O_{14}}$}

\begin{figure*} 
	\centering\includegraphics[scale=0.62]{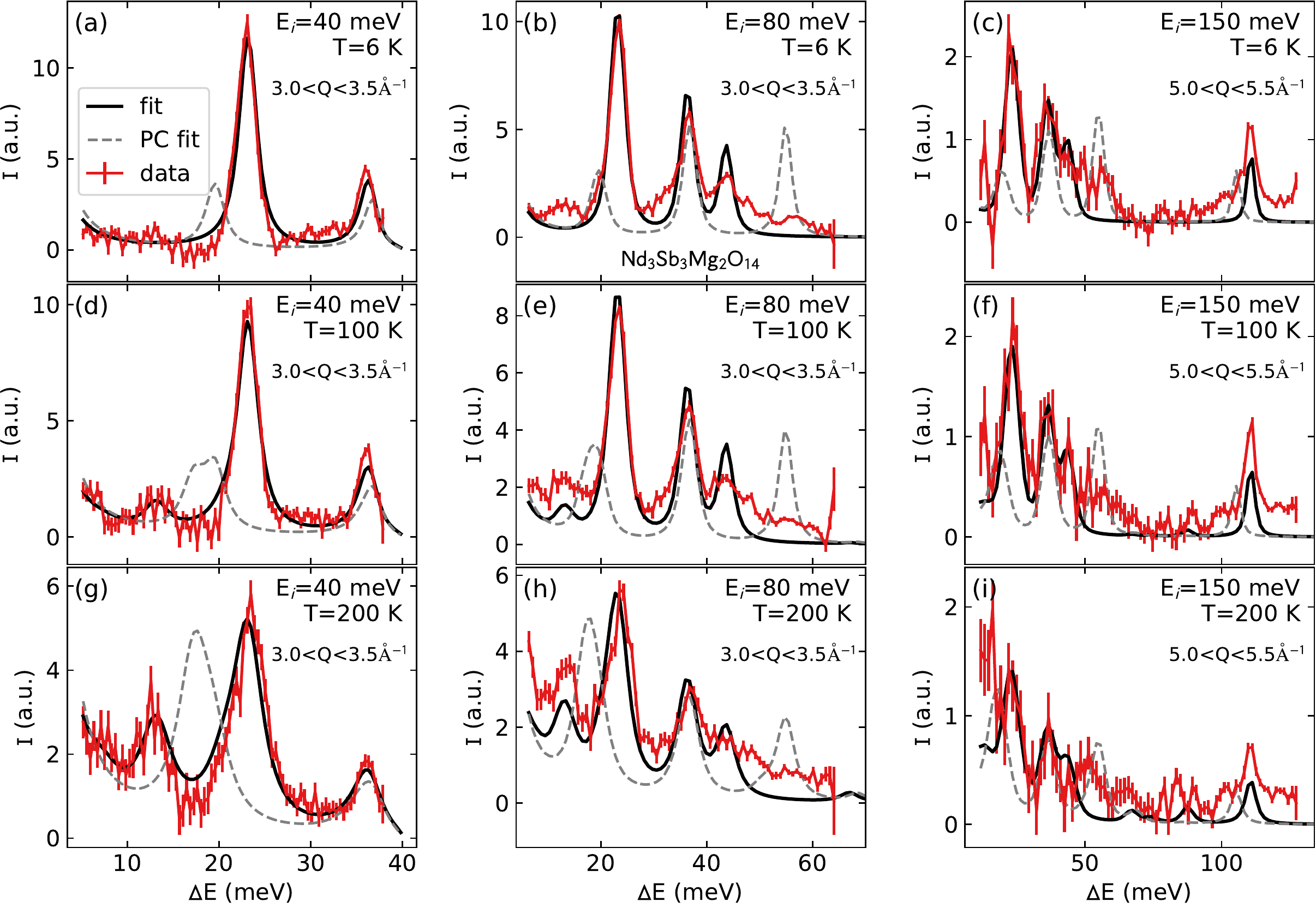}
	
	\caption{Constant Q cuts showing the results of the CEF fit to  $\rm{Nd_3Sb_3Mg_2O_{14}}$ neutron scattering data.
		The point charge fit ("PC fit") is shown with a grey dashed line, and the final fit ("fit") is shown with a solid black line.}
	\label{flo:NdMg_Qcuts}
\end{figure*}

The best fit CEF parameters for $\rm{Nd_3Sb_3Mg_2O_{14}}$ are listed in Table \ref{flo:NdMg_CEF_params}, along with the  CEF parameters from the initial Calculated PC and the PC Fit models. Constant Q cuts of the fits to neutron data are shown in Fig. \ref{flo:NdMg_Qcuts}, with the final fit plotted in black and the PC fit plotted in a grey dashed line.

\begin{table}
	\caption{Calculated and fitted CEF parameters for $\rm{Nd_3Sb_3Mg_2O_{14}}$. The first column (Calculated PC) is the CEF parameters from a point-charge model where all effective charges are 2$e$. The second column (PC Fit) gives the result of the effective charge fit. The final column (Final Fit) is the result of fitting the CEF parameters to the data.}
	\begin{ruledtabular}
		\begin{tabular}{c|ccc}
			$B_n^m$ (meV) &Calculated PC & PC Fit & Final Fit \tabularnewline
			\hline 
$ B_2^0$ & 0.08051 & -0.1851 & 0.0121 \tabularnewline
$ B_2^1$ & -0.5358 & -0.80854 & -0.25649 \tabularnewline
$ B_2^2$ & 0.04892 & -0.00787 & -0.02649 \tabularnewline
$ B_4^0$ & -0.0131 & -0.01914 & -0.01861 \tabularnewline
$ B_4^1$ & 0.00181 & 0.00297 & 0.00844 \tabularnewline
$ B_4^2$ & -0.00339 & -0.00441 & 0.00763 \tabularnewline
$ B_4^3$ & -0.11134 & -0.15368 & -0.04106 \tabularnewline
$ B_4^4$ & 0.00772 & 0.00994 & 0.0198 \tabularnewline
$ B_6^0$ & -0.00018 & -0.00027 & -0.00056 \tabularnewline
$ B_6^1$ & 3$\times 10^{-5}$ & 3$\times 10^{-5}$ & 0.00011 \tabularnewline
$ B_6^2$ & 0.00017 & 0.00023 & -0.00028 \tabularnewline
$ B_6^3$ & 0.00212 & 0.00293 & -0.00138 \tabularnewline
$ B_6^4$ & -0.00023 & -0.00031 & -0.00052 \tabularnewline
$ B_6^5$ & -0.00055 & -0.00081 & -0.00073 \tabularnewline
$ B_6^6$ & -0.00224 & -0.00309 & -0.00218 \tabularnewline
		\end{tabular}\end{ruledtabular}
		\label{flo:NdMg_CEF_params}
	\end{table}

Starting with effective charges of ($-2e$, $-2e$, $-2e$), the PC fitted charges for $\rm{Nd_3Sb_3Mg_2O_{14}}$ are ($-0.999e$,  $-0.931e$,  $-0.910e$). The Powell method of minimization \cite{PowellsMethod} yields this result for any value of initial charges from $-0.6e$ to $-2e$. Although these values are about 50\% less than $-2e$, they are reasonable because the electrostatic repulsion is actually from electron orbitals and not point charges; so the effective charge can differ significantly from the net charge \cite{newman2007crystal}. As Fig. \ref{flo:NdMg_Qcuts} shows, the effective charge fit resembles the data but does not reproduce the precise energies and intensities of the transitions. The final fit matches the data much better, with the location and intensity of all major peaks reproduced. 

The ground state eigenstates from the Calculated PC and the Final Fit are listed in Table \ref{flo:NdMgEigenvectors}. In both fits the ground state doublet is mostly $|\pm \frac{9}{2}\rangle$, with some weight given to $|\pm \frac{3}{2}\rangle$. For the complete set of eigenkets, see the supplemental materials.

The ground state ordered moment, computed from $\langle 0 | J_{\alpha} | 0 \rangle$ is $\langle J_{x} \rangle = \pm 0.11 \> \mu_B$, $\langle J_{y} \rangle = 0.00 \> \mu_B$, $\langle J_{z} \rangle = \mp 2.89 \> \mu_B$, for a total $\langle J \rangle = \sqrt{\sum_{\alpha} \langle J_{\alpha} \rangle^2 } =  2.89 \> \mu_B$.

\begin{table*}
	\caption{Ground state eigenvectors and eigenvalues for $\rm{Nd_3Sb_3Mg_2O_{14}}$. The top two lines give the results of the Calculated PC model, and the last two lines give the results of the Final Fit. In both cases the ground state kets are primarily $| \pm 9/2 \rangle$.}
	\begin{ruledtabular}
		\begin{tabular}{c||c|cccccccccc}
			Model &E (meV) &$| -\frac{9}{2}\rangle$ & $| -\frac{7}{2}\rangle$ & $| -\frac{5}{2}\rangle$ & $| -\frac{3}{2}\rangle$ & $| -\frac{1}{2}\rangle$ & $| \frac{1}{2}\rangle$ & $| \frac{3}{2}\rangle$ & $| \frac{5}{2}\rangle$ & $| \frac{7}{2}\rangle$ & $| \frac{9}{2}\rangle$ \tabularnewline
			\hline 
PC calc. & 0.000 & 0.8181 & -0.0632 & -0.0772 & -0.1835 & 0.1644 & 0.1965 & 0.4597 & 0.0936 & 0.036 & -0.0064 \tabularnewline
& 0.000 & 0.0064 & 0.036 & -0.0936 & 0.4597 & -0.1965 & 0.1644 & 0.1835 & -0.0772 & 0.0632 & 0.8181 \tabularnewline
			\hline 
Final Fit & 0.000 & 0.8346 & 0.0211 & -0.0939 & -0.291 & 0.0711 & -0.0357 & 0.4097 & 0.0782 & -0.0248 & 0.1693 \tabularnewline
& 0.000 & 0.1693 & 0.0248 & 0.0782 & -0.4097 & -0.0357 & -0.0711 & -0.291 & 0.0939 & 0.0211 & -0.8346 \tabularnewline
		\end{tabular}\end{ruledtabular}
		\label{flo:NdMgEigenvectors}
	\end{table*}

\subsection{$\rm{Nd_3Sb_3Zn_2O_{14}}$}

\begin{figure*} 
	\centering\includegraphics[scale=0.62]{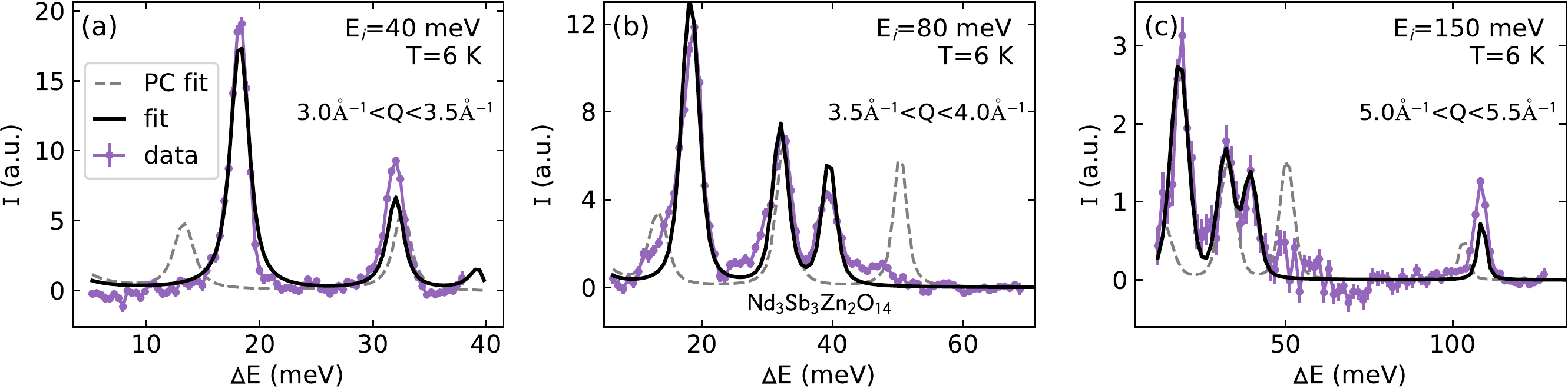}
	
	\caption{Constant Q cuts showing the results of the CEF fit to  $\rm{Nd_3Sb_3Zn_2O_{14}}$ neutron scattering data. Only $T=6$ K data is shown; 100 K and 200 data are shown in the Supplemental Materials.
	}
	\label{flo:NdZn_Qcuts}
\end{figure*}

The results of the fits to $\rm{Nd_3Sb_3Zn_2O_{14}}$ data are similar to $\rm{Nd_3Sb_3Mg_2O_{14}}$.
Constant Q cuts of the Final Fit to  $\rm{Nd_3Sb_3Zn_2O_{14}}$ CEF neutron data are shown in Fig. \ref{flo:NdZn_Qcuts}. The effective charge fit (PC Fit) yielded ($-1.01e$, $-0.968e$, $-0.915e$). The PC Fit resembles the data, but the final fit matches the data much better and provides  a faithful reproduction of all large peaks.

The ground state eigenkets from the final fit are listed in Table \ref{flo:NdZnEigenvectors}. Like  $\rm{Nd_3Sb_3Mg_2O_{14}}$, the ground state doublet is mostly composed of $|\pm \frac{9}{2}\rangle$, with some $|\pm \frac{3}{2}\rangle$ also present. The initial point-charge calculation (Calculated PC) predicted significant weight on $|\pm \frac{1}{2}\rangle$ which is not present in the final fit.  This indicates that the point-charge model, while it is a good starting point for fits, does not reliably predict the nature of the ground state doublet for these low-symmetry ligand environments.  Plots of Q-cuts of higher temperature data, the list of fitted CEF parameter values, and a full list of eigenstates can be found in the supplemental materials.

The ground state ordered moment, computed from $\langle 0 | J_{\alpha} | 0 \rangle$ is $\langle J_{x} \rangle = \pm 0.23  \> \mu_B$, $\langle J_{y} \rangle = 0  \> \mu_B$, $\langle J_{z} \rangle = \mp 2.40  \> \mu_B$. The total ordered moment of $\langle J \rangle = 2.41  \> \mu_B$  is slightly less than for $\rm{Nd_3Sb_3Mg_2O_{14}}$.

\begin{table*}
	\caption{Ground state wavefunctions for $\rm{Nd_3Sb_3Zn_2O_{14}}$. The top two lines give the results from the Calculated PC model, and the last two lines give the results of the final fit. In this case, the point-charge model involves $| \pm 1/2 \rangle$ while the final fit shifts most of the weight to $| \pm 9/2 \rangle$.}
	\begin{ruledtabular}
		\begin{tabular}{c||c|cccccccccc}
			Model &E (meV) &$| -\frac{9}{2}\rangle$ & $| -\frac{7}{2}\rangle$ & $| -\frac{5}{2}\rangle$ & $| -\frac{3}{2}\rangle$ & $| -\frac{1}{2}\rangle$ & $| \frac{1}{2}\rangle$ & $| \frac{3}{2}\rangle$ & $| \frac{5}{2}\rangle$ & $| \frac{7}{2}\rangle$ & $| \frac{9}{2}\rangle$ \tabularnewline
			\hline 
PC calc. &0.000 & 0.4198 & -0.0573 & -0.2189 & 0.1273 & -0.4703 & -0.5831 & 0.4015 & 0.1527 & 0.1031 & 0.0 \tabularnewline
&0.000 & 0.0 & -0.1031 & 0.1527 & -0.4015 & -0.5831 & 0.4703 & 0.1273 & 0.2189 & -0.0573 & -0.4198 \tabularnewline
			\hline 
Final Fit &0.000 & 0.2368 & 0.0265 & 0.0455 & 0.4932 & 0.0389 & 0.0939 & 0.1583 & 0.0049 & -0.0336 & 0.8133 \tabularnewline
&0.000 & 0.8133 & 0.0336 & 0.0049 & -0.1583 & 0.0939 & -0.0389 & 0.4932 & -0.0455 & 0.0265 & -0.2368 \tabularnewline
		\end{tabular}\end{ruledtabular}
		\label{flo:NdZnEigenvectors}
	\end{table*}

	

\subsection{$\rm{Pr_3Sb_3Mg_2O_{14}}$}

\begin{figure*} 
	\centering\includegraphics[scale=0.62]{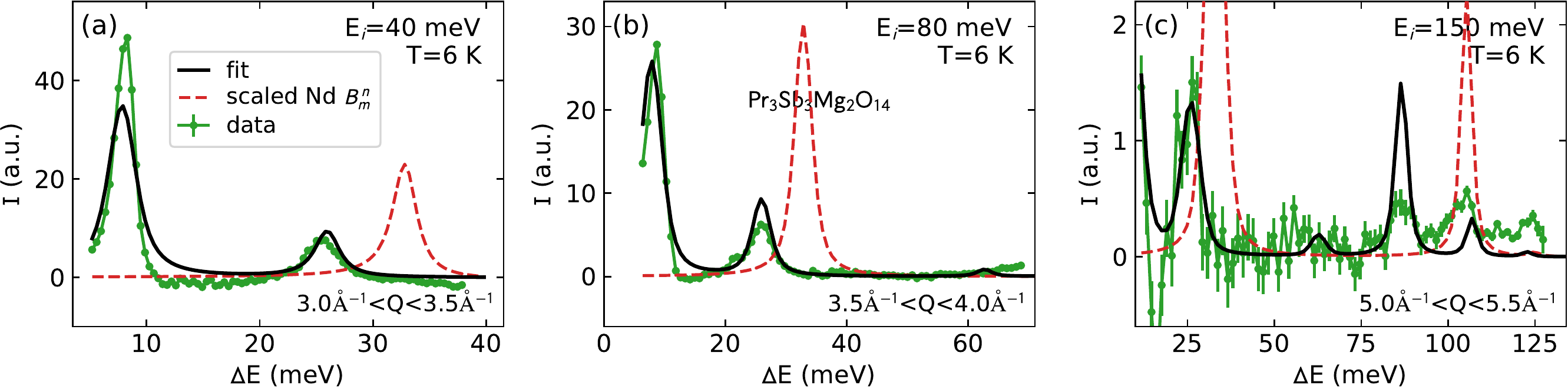}
	
	\caption{Constant Q cuts showing the results of the final fit to  $\rm{Pr_3Sb_3Mg_2O_{14}}$ neutron scattering data, along with the results of rescaling the CEF parameters from $\rm{Nd_3Sb_3Mg_2O_{14}}$. Only $T=6$ K data are shown; 100 K and 200 data are shown in the Supplemental Materials. Clearly, the rescaled CEF parameters are unreliable. The black fit line is not as good as for the Nd compounds, especially for the higher energy transitions in panel (c).
	}
	\label{flo:PrMg_Qcuts}
\end{figure*}

Constant Q cuts of the final fit to  $\rm{Pr_3Sb_3Mg_2O_{14}}$ CEF neutron data are shown in Fig. \ref{flo:PrMg_Qcuts}.  Because Pr$^{3+}$ is a non-Kramers ion, non-magnetic singlets are possible and there are many more energy levels and transitions. An unfortunate consequence of this is that many of the transitions are too faint to distinguish, and the neutron spectrum fit is mostly based on the low energy ($<50$ meV) data. Accordingly, the $\chi^2_{\Delta E}$ term for $\rm{Pr_3Sb_3Mg_2O_{14}}$ only gave significant weight to the lowest two observed energies. The PC Fit charges from the effective point charge model are ($-0.805e$  $-0.736e$,  $-0.836e$). The lowest two eigenstates and eigenkets from the final fit are listed in Table \ref{flo:PrMgEigenvectors}. As required by group theory, $\langle j_x \rangle = \langle j_y \rangle = \langle j_z \rangle =0$ for all singlet states. Plots of Q-cuts of higher temperature data, the list of fitted CEF parameter values, and a full list of eigenstates can be found in the supplemental materials.

The final fit resembles the data reasonably well (Fig. \ref{flo:PrMg_Qcuts}), with the exception of a predicted peak at 60 meV and too much intensity on the 85 meV peak. Nevertheless, the final fit appears to be close.

\begin{table*}
	\caption{Eigenvectors and eigenvalues for the ground state and first excited crystal field state of  $\rm{Pr_3Sb_3Mg_2O_{14}}$. The top two lines give the results of the calculated PC model, and the last two lines give the results of the final fit. As a non-Kramers ion, the ground state is not constrained to be a doublet. As required for singlets, $\langle j_x \rangle = \langle j_y \rangle = \langle j_z \rangle =0$ for all states.}
	\begin{ruledtabular}
		\begin{tabular}{c||c|ccccccccc}
			Model & E (meV) &$|-4\rangle$ & $|-3\rangle$ & $|-2\rangle$ & $|-1\rangle$ & $|0\rangle$ & $|1\rangle$ & $|2\rangle$ & $|3\rangle$ & $|4\rangle$ \tabularnewline
			\hline 
		PC calc. &0.000 & 0.0211 & 0.1351 & 0.0143 & 0.0677 & -0.9762 & -0.0677 & 0.0143 & -0.1351 & 0.0211 \tabularnewline
		&9.028 & 0.481 & -0.0515 & -0.0064 & -0.5137 & -0.0648 & 0.5137 & -0.0064 & 0.0515 & 0.481 \tabularnewline
			\hline 
Final Fit &0.000 & 0.2851 & -0.1002 & 0.2334 & 0.3867 & -0.6398 & -0.3867 & 0.2334 & 0.1002 & 0.2851 \tabularnewline
&7.963 & -0.1683 & -0.0653 & 0.0903 & -0.4922 & -0.6587 & 0.4922 & 0.0903 & 0.0653 & -0.1683 \tabularnewline
		\end{tabular}\end{ruledtabular}
		\label{flo:PrMgEigenvectors}
	\end{table*}

An alternative to directly fitting a CEF model is re-scaling the CEF parameters from a compound with a similar ligand environment. We carried out such a calculation for $\rm{Pr_3Sb_3Mg_2O_{14}}$ by re-scaling the CEF parameters from the final fit $B_n^m$ from $\rm{Nd_3Sb_3Mg_2O_{14}}$ using the equation
\begin{equation}
\big( B_{n}^{m} \big)_{\rm Pr} = \big( B_{n}^{m} \big)_{\rm Nd}
\frac{\left\langle r^{n}\right\rangle_{\rm Pr} \theta_{n\> {\rm Pr}}}{\left\langle r^{n}\right\rangle_{\rm Nd} \theta_{n \> {\rm Nd}}},
\end{equation}
which is derived from Eq. \ref{eq:CEFparams} for two different ions with the same ligand environment. While the ligand environments are not identical, this re-scaling sometimes works for two rare earth ions with similar electron counts \cite{Carnall1989}. The results are plotted in Fig. \ref{flo:PrMg_Qcuts}. Unfortunately, the re-scaled CEF parameters do not come close to predicting the energy or intensity of the transitions in the neutron spectrum. Therefore, we conclude that it is not possible to rescale the CEF parameters to accurately predict the CEF Hamiltonians of $\rm{RE_3Sb_3A_2O_{14}}$.

\subsection{Susceptibility}

\begin{figure} 
	\centering\includegraphics[scale=0.5]{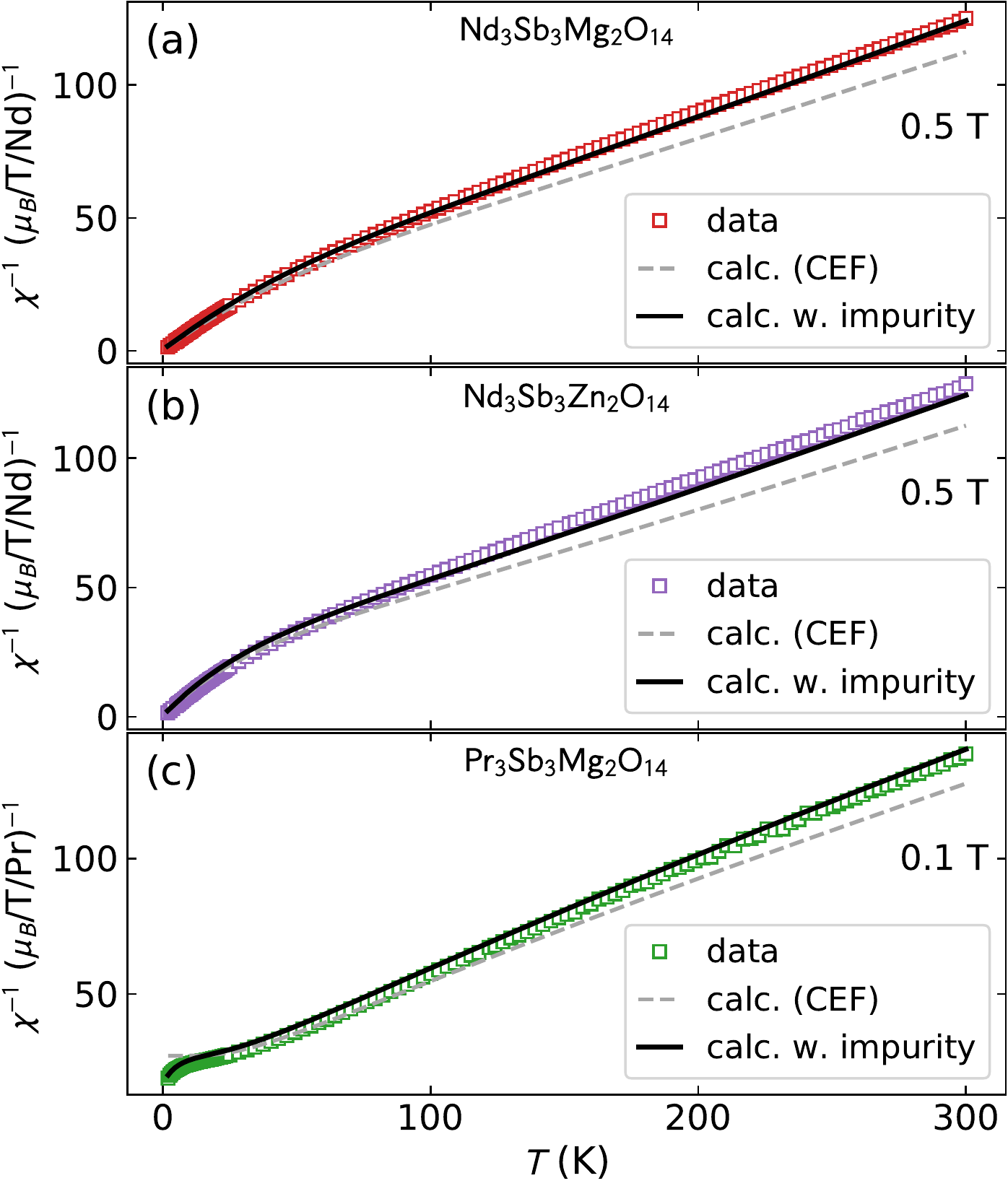}
	
	\caption{Comparison between the measured susceptibility and the susceptibility calculated from the final fit CEF Hamiltonians. In each case the calculation overestimates the high temperature susceptibility by about 10\%, which may result from chemical impurities in the samples used for susceptibility measurements. 
	}
	\label{flo:Susceptibility}
\end{figure}

The calculated magnetic susceptibilities for all three compounds based on the final fit CEF Hamiltonians are plotted in Fig. \ref{flo:Susceptibility}, along with experimental data from refs. \cite{MyPaper,SandersREMg,SandersREZn}. In every case, the $\chi_{CEF}$ calculation (plotted with a gray dashed line) overestimates the measured susceptibility by about 10\% (the predicted inverse susceptibility curve lies below the data). The reason for this discrepancy appears to be impurities or site-mixing in the compounds, mainly evidenced by the low-temperature $\rm{Pr_3Sb_3Mg_2O_{14}}$ data.

In the $\rm{Pr_3Sb_3Mg_2O_{14}}$ experimental susceptibility plotted in Fig. \ref{flo:Susceptibility}(c), $\chi^{-1} \rightarrow 0$ as $T \rightarrow 0$. This should not happen for a singlet ground state (non-Kramers ion in a low ligand field), where $\chi$ should saturate at a finite value. The deviation to zero indicates Kramers ions in the sample. To estimate the relative contribution, we fit the $\rm{Pr_3Sb_3Mg_2O_{14}}$ susceptibility to a simple model: $\chi = (x) \> \chi_{CEF} + (1-x) \chi_{CW}$, where $0 < x < 1$ and $\chi_{CW}$ is represented by a Curie-Weiss law: $\frac{C}{T-\theta_C}$. The fit works surprisingly well, and indicates a 13\% orphan spin contribution with an effective moment of 1.8~$\mu_B$ (see Supplemental Information for more details). Such a contribution could arise from site mixing between Pr and Mg, like the $\sim 10 \%$ Dy/Mg site mixing observed in $\rm{Dy_3Sb_3Mg_2O_{14}}$ \cite{Paddison2016}. This would decouple some of the spins from the kagome planes, and put them in completely different ligand environments.

We also attempted to account for the susceptibility discrepancy using an interaction model $\chi = \frac{ \chi_{CEF}}{1- \lambda \chi_{CEF}}$ where $\lambda$ is the magnetic interaction between ions. No matter what $\lambda$ is chosen, model fails to account for the low temperature divergence, and it fails to correct the slope of high temperature susceptibility. Thus, the observed effects indicate an additional Cure-Weiss contribution to the susceptibility and not merely interactions.

Incorporating this Curie-Weiss contribution model makes the calculations match the low-temperature $\rm{Pr_3Sb_3Mg_2O_{14}}$ susceptibility data well, and happens to resolve the high-temperature discrepancy between theory and experiment. Assuming that the $\rm Nd^{3+}$ compounds have the same $\chi_{CW}$, we also get good agreement between theory and experiment for $\rm{Nd_3Sb_3Mg_2O_{14}}$ and $\rm{Nd_3Sb_3Zn_2O_{14}}$ (Fig. \ref{flo:Susceptibility}).

We tested and ultimately rejected three alternative explanations for the high-temperature discrepancy between calculated and measured susceptibility: (i) an incorrect CEF Hamiltonian, (ii) sample diamagnetism and (iii) higher multiplet mixing. We tested (i) by attempting to re-fit the CEF Hamiltonian to the neutron data including a $\chi^2$ term from calculated susceptibility (without $\chi_{CW}$). This attempt failed. No matter what starting parameters are chosen (and the relative $\chi^2$ weight given to susceptibility versus neutron spectrum), we were unable to fit them simultaneously. 
We tested (ii) by measuring the susceptibility of the nonmagnetic analogue $\rm{La_3Sb_3Mg_2O_{14}}$, which comes out to  $-10^{-4}$ ($\mu_B$/T/ion)---an order of magnitude too small. We tested (iii) by calculating susceptibility using the intermediate coupling-scheme and found that the result is nearly identical to the fits based on the Hunds rule spin-orbital ground state. (Details behind (ii) and (iii) are given in the Supplemental Information.) Therefore, we are confident that the discrepancy between calculated and measured susceptibility in $\rm{Pr_3Sb_3Mg_2O_{14}}$ is due to  orphan Kramers ions in the sample.

10\% population of orphan spins is too little to detect and significantly affect the CEF excitation spectrum. However, it may be enough to have significant effects on some forms of collective phenomena in these frustrated magnets. 

\section{Discussion}

The point-charge fit followed by the final CEF parameter fit seems to have worked as a method to determine the crystal field level scheme in the low point group symmetry $\rm{RE_3Sb_3A_2O_{14}}$ compounds. The final fit matches the data well, and along the way the fitted effective charges are within an electron charge from the formal ligand charge. Furthermore the calculated temperature dependent susceptibility reproduces measurements well after accounting for orphan spins at the <10\% level. We are confident that we have identified the single-ion CEF Hamiltonians for $\rm{Nd_3Sb_3Mg_2O_{14}}$,  $\rm{Nd_3Sb_3Zn_2O_{14}}$, and $\rm{Pr_3Sb_3Mg_2O_{14}}$ and determined the associated crystal field eigenvalues and eigenstates.

The analysis of $\rm{Nd_3Sb_3Zn_2O_{14}}$ shows that the point-charge model by itself does not reliably predict the ground state eigenkets of $\rm{RE_3Sb_3A_2O_{14}}$ compounds. Here we note that we are basing the models on a high T x-ray structural refinement. Low T neutron diffraction measurements would provide more accurate ligand positions, which could improve the point charge fitting.  We find that scaling $\rm Nd^{3+}$ results to $\rm Pr^{3+}$ does not reproduce the observed spectrum in $\rm{Pr_3Sb_3Mg_2O_{14}}$. Therefore, it is unfortunately not possible to accurately  predict the CEF ground states of other $\rm{RE_3Sb_3A_2O_{14}}$ compounds from these results.

For $\rm Pr^{3+}$ the lowest level states of the single-ion non-Kramers states are singlets. This is true for the naive calculated PC Hamiltonian, the PC fit Hamiltonian, and the final fit Hamiltonian. The gap between the lowest and first excited state exceeds the exchange energy scale so that we expect this system to be a singlet ground state system with no phase transitions. 

\begin{figure} 
	\centering\includegraphics[scale=0.55]{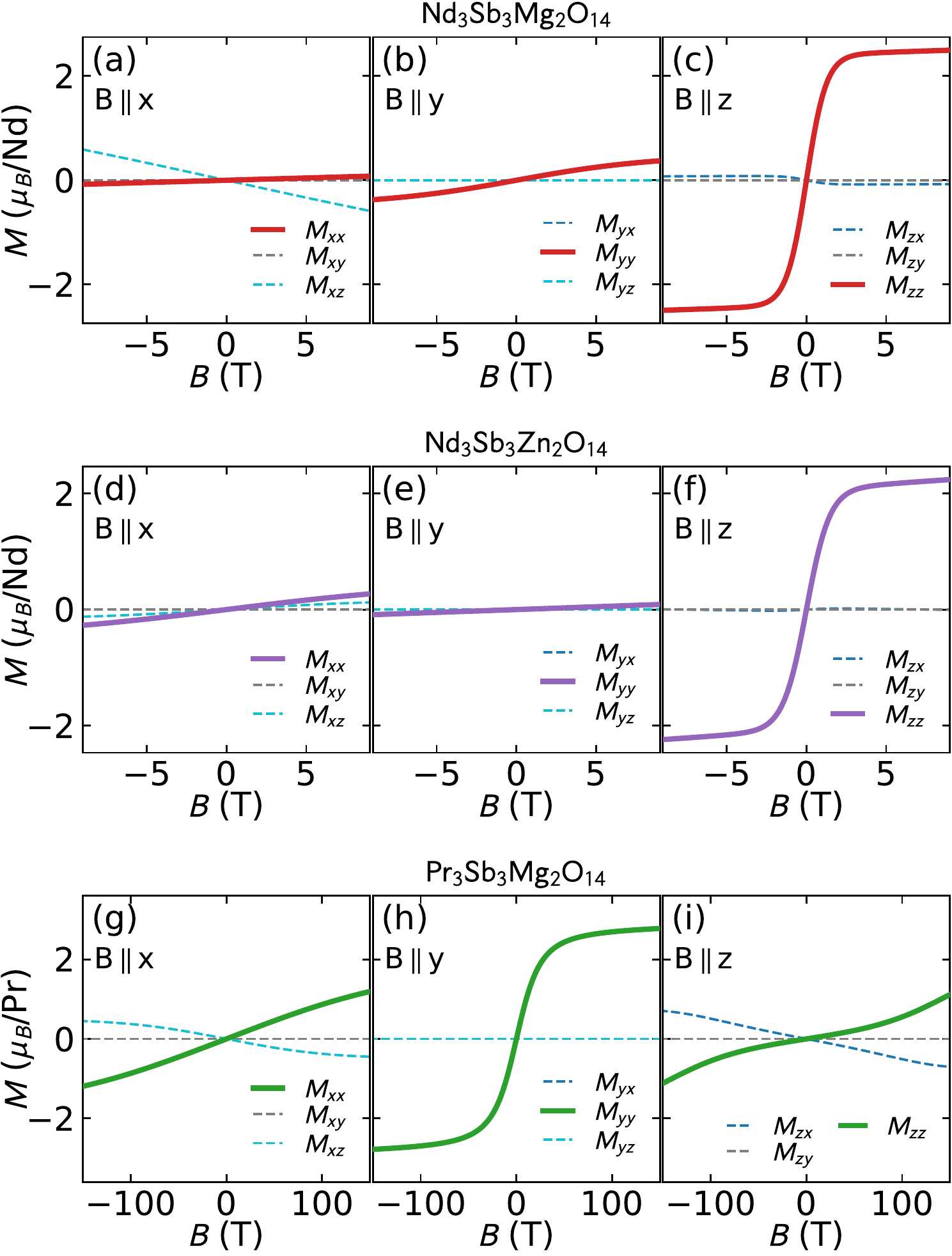}
	
	\caption{Directional single ion magnetization computed from the final fit CEF Hamiltonians for $\rm{Nd_3Sb_3Mg_2O_{14}}$,  $\rm{Nd_3Sb_3Zn_2O_{14}}$, and  $\rm{Pr_3Sb_3Mg_2O_{14}}$ at 2 K. The directions $x$, $y$, and $z$ are defined in Fig. \ref{flo:LigandEnvt}.
	}
	\label{flo:Anisotropy}
\end{figure}

One of the key features of interest for these compounds is the magnetic anisotropy. One can gain a rough understanding of the single ion anisotropy by examining the ground state wave function. The final fit results for $\rm{Nd_3Sb_3Mg_2O_{14}}$ and $\rm{Nd_3Sb_3Zn_2O_{14}}$ have mostly an effective $J=|\pm \frac{9}{2}\rangle$ ground state doublet, which can be interpreted as easy-axis moments. The substitution of Zn for Mg does not have a dramatic effect on the ground state, at least for the $\rm Nd^{3+}$ ion. 
For a clearer picture of the anisotropy, the computed single-ion directional magnetization at 2 K is shown in Fig. \ref{flo:Anisotropy}. For $\rm{Nd_3Sb_3Mg_2O_{14}}$ and $\rm{Nd_3Sb_3Zn_2O_{14}}$, the saturation magnetic field is around 5 T, with the largest magnetization for $B \parallel z$, indicating an easy-axis. Negligibly small off-diagonal elements exist for the $x$ and $z$ directions. For $\rm{Pr_3Sb_3Mg_2O_{14}}$, the predicted saturation magnetic field is around 80 T with the easiest axis in the $y$ direction.

These results show that the authors' previously hypothesized effective $J=|\pm \frac{1}{2}\rangle$ Nd$^{3+}$ ground state for $\rm{Nd_3Sb_3Mg_2O_{14}}$ \cite{MyPaper} is incorrect, and Dun et. al.'s suggestion of an easy axis \cite{Dun2017} is closer to the true ground state. Ref. \cite{MyPaper} failed to account for impurities in magnetization, which led to the inference of an incorrect model.

The ordered moment in $\rm{Nd_3Sb_3Mg_2O_{14}}$ determined from neutron scattering is $1.79(5)\> \mu_B$ \cite{MyPaper}. Assuming a 13\% site-mixing, this is only 71\% of the theoretically predicted moment of $(0.87\times 2.89\> \mu_B) = 2.51\> \mu_B $. This reduction in moment, in conjunction with the magnetic entropy not reaching $R\ln(2)$ \cite{MyPaper}, suggests that the magnetism in $\rm{Nd_3Sb_3Mg_2O_{14}}$ remains dynamic to the lowest temperatures. This suggests a closer examination of the collective properties of this material in a high quality single crystal sample would be interesting.

\section{Conclusion}

We have outlined a method whereby complex inelastic neutron scattering spectra for crystal field excitations of rare earth ions can be fitted using a point-charge model with effective ligand charges as an intermediate step. We applied this method to Nd$^{3+}$ in $\rm{Nd_3Sb_3Mg_2O_{14}}$ and $\rm{Nd_3Sb_3Zn_2O_{14}}$, showing that the single-ion anisotropy is easy-axis. We also applied the method to Pr$^{3+}$, showing that the ground state is a singlet with an energy gap of 8.0 meV.

This information is an essential component towards understanding the low-temperature magnetism of this new family of frustrated magnets, and will guide further investigations of their collective properties. 

\paragraph{Note:} While this manuscript was in the final stages of preparation, there appeared Ref. \cite{dun2018quantum} which independently implemented an effective point charge fit to the CEF Hamiltonian of $\rm Ho_3Sb_3Mg_2O_{14}$.

\section*{Acknowledgments}
This work was supported through the Institute for Quantum Matter at Johns Hopkins University, by the U.S. Department of Energy, Division of Basic Energy Sciences, Grant DE-FG02-08ER46544. AS and CB were supported through the Gordon and Betty Moore foundation under the EPIQS program GBMF4532. ADC was partially supported by the U.S. DOE, Office of Science, Basic Energy Sciences, Materials Sciences and Engineering Division. This research at the High Flux Isotope Reactor and Spallation Neutron Source was supported by DOE Office of Science User Facilities Division. AS acknowledges helpful discussions with Andrew Boothroyd.

\newpage
\vfill
~
\newpage

\section*{Supplemental Material}

\renewcommand{\thefigure}{S\arabic{figure}}
\renewcommand{\theequation}{S.\arabic{equation}}
\renewcommand{\thepage}{S\arabic{page}}  
\renewcommand{\thesection}{S.\Roman{section}}
\renewcommand{\thetable}{S.\Roman{table}}

\setcounter{figure}{0}
\setcounter{page}{1}
\setcounter{equation}{0}
\setcounter{section}{0}
\setcounter{table}{0}

\section{Experimental Methods}

The following section describes the background subtraction and Debye-Waller fit to the neutron scattering data.

\subsection{Background Subtraction}

To isolate the magnetic signal in the neutron spectrum for $\rm{Nd_3Sb_3Mg_2O_{14}}$,  $\rm{Nd_3Sb_3Zn_2O_{14}}$, and  $\rm{Pr_3Sb_3Mg_2O_{14}}$, we measured and subtracted the scattering from nonmagnetic analogue $\rm{La_3Sb_3Mg_2O_{14}}$. The structure of $\rm{La_3Sb_3Mg_2O_{14}}$ being very similar to its magnetic counterparts, the phonon spectrum should be nearly identical. Scaling the intensity (which is different because of different cross sections of the substituted atoms) should account for the differences.

\begin{figure}
	\centering\includegraphics[scale=0.65]{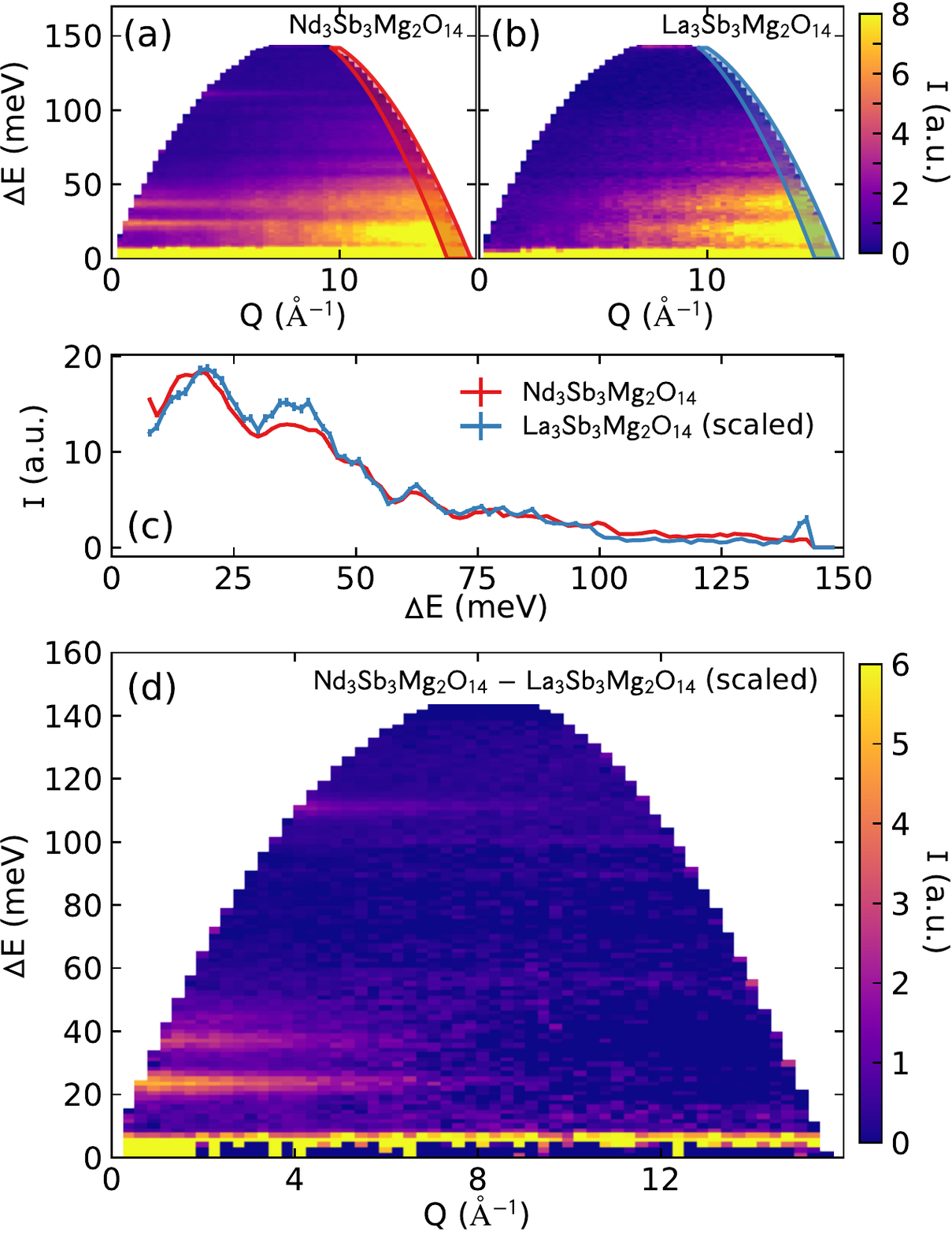}
	
	\caption{Background subtraction for the neutron data. (a) Raw neutron scattering data for $\rm{Nd_3Sb_3Mg_2O_{14}}$ at $E_i=150$ meV and 6 K. (b) Raw neutron scattering data for nonmangetic $\rm{La_3Sb_3Mg_2O_{14}}$ at $E_i=150$ meV and 6 K. The red and blue bars on the right of the panels demarcate $120^{\circ} < 2\theta < 136^{\circ}$, which are plotted vs $\Delta E$ in panel (c) with the La compound intensity scaled to match the Nd compound intensity. (d) shows the results of subtracting the La compound intensity from the Nd compound intensity.
	}
	\label{flo:BGsubtraction}
\end{figure}

To avoid magnetic signals influencing the background scaling, we only scaled the high-Q scattering---where the form factor should suppress the magnetic signal. For every $E_i$ and temperature we scaled the La nonmagnetic analogue by minimizing the $\chi^2$ difference between the magnetic and nonmangetic scattering in a cut through $120^{\circ} < 2\theta < 136^{\circ}$, as shown in Fig. \ref{flo:BGsubtraction}. 

We measured the spectrum of $\rm{La_3Sb_3Mg_2O_{14}}$ at $E_i=150$ meV, 80 meV, and 40 meV and at 200 K, 100K, and 6 K (the same configurations for the data). $\rm{La_3Sb_3Mg_2O_{14}}$ scattering was measured with different shutter settings than the rest of the compounds, so we are unable to directly compare the scaling factors with the expected cross-section ratios. Nevertheless, the background-subtracted data sets [for example, Fig. \ref{flo:BGsubtraction}(d)] reveal the electronic crystal-field signals very clearly.

\subsection{Debye Waller Factor Fit}

\begin{figure*}
	\centering\includegraphics[scale=0.56]{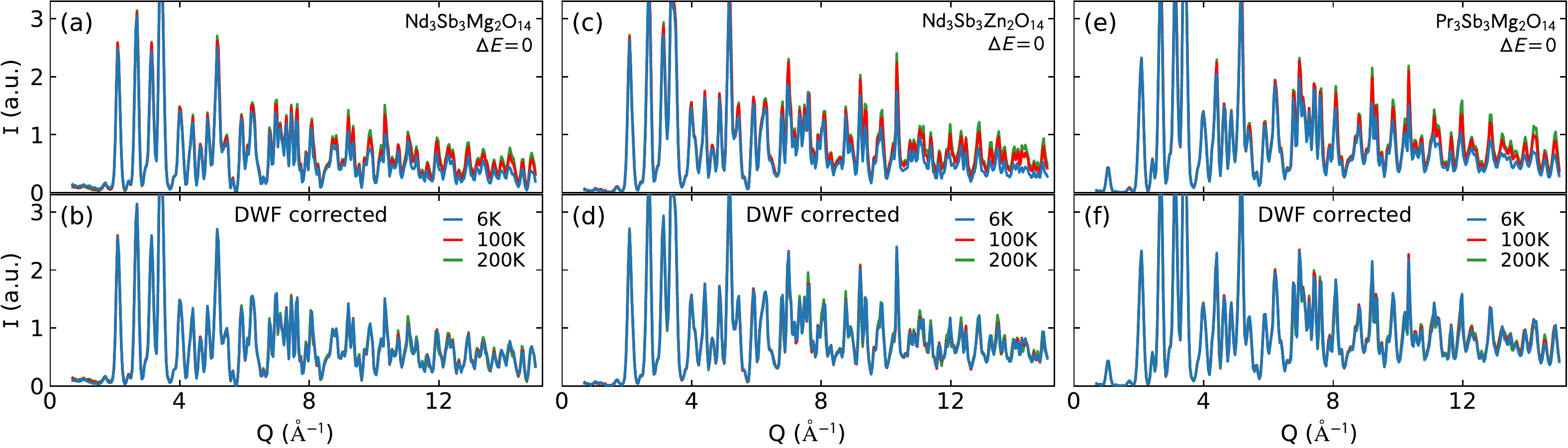}
	
	\caption{Debye Waller Factor fits for $\rm{Nd_3Sb_3Mg_2O_{14}}$ (a)-(b),  $\rm{Nd_3Sb_3Zn_2O_{14}}$ (c)-(d), and  $\rm{Pr_3Sb_3Mg_2O_{14}}$ (e)-(f) based on elastic scattering for 6 K (blue), 100 K (red), and 200 K (green). The top panels show the raw data and the bottom panels show the 100 K and 200 K data multiplied by $\exp(\frac{1}{3}Q^2 \langle u^2 \rangle)$, where $\langle u^2 \rangle$ was fitted by minimizing the difference with the 6 K data.
	}
	\label{flo:DWF_fit}
\end{figure*}

The neutron scattering of finite-temperature materials is modulated by $e^{2W ({\bf Q})}$, found in eq. 3 of the main text, where $W ({\bf Q})$ is the Debye Waller factor which arises from thermal vibrations of an atom about its average position. We can write this factor as 
\begin{equation}
2 W ({\bf Q}) = \frac{1}{3} {\bf Q}^2 \langle u^2 \rangle
\end{equation}
where $\langle u^2 \rangle$ is the average displacement of the magnetic ion at a given temperature \cite{Squires}. To estimate $\langle u^2 \rangle$, we assume that the DW factor is negligible at 6K, and find the value of $\langle u^2 \rangle$ necessary to make the 100 K and 200 K elastic data match the 6 K elastic data. This approach is an approximation because it assumes the same $\langle u^2 \rangle$ for all atoms; but it works reasonably well in describing the Q dependence of the scattering (see Fig. \ref{flo:DWF_fit}).

We fit $\langle u^2 \rangle$ for each temperature by minimizing the $\chi^2$ difference between the higher temperature scattering and the 6 K elastic scattering, fitting the $E_i=150$ meV, 80 meV, and 40 meV data simultaneously for each temperature and each compound. Based on the resolution function defined for ARCS, we took the elastic scattering to be $\pm 3$ meV for $E_i=150$ meV, $\pm 1.6$ meV for $E_i=80$ meV, and $\pm 1.3$ meV for $E_i=40$ meV. The elastic intensities before and after scaling for $E_i=150$ meV are shown in Fig. \ref{flo:DWF_fit}, and the fitted values of $\langle u^2 \rangle$ are shown in table \ref{flo:Fitted_DWF}. As expected, $\langle u^2 \rangle$ varies roughly linearly with temperature (the relationship for a simple harmonic oscillator).

\begin{table}
	\caption{Fitted $\langle u^2 \rangle$ for the Debye-Waller factor.}
	\begin{ruledtabular}
		\begin{tabular}{c|ccc}
			$T$ (K) &$\rm{Nd_3Sb_3Mg_2O_{14}}$ &  $\rm{Nd_3Sb_3Zn_2O_{14}}$ &  $\rm{Pr_3Sb_3Mg_2O_{14}}$ \tabularnewline
			\hline 
			100 & 0.0451 & 0.0452 & 0.0459 \tabularnewline
			200 & 0.0849 & 0.0878 & 0.0769 \tabularnewline
		\end{tabular}\end{ruledtabular}
		\label{flo:Fitted_DWF}
	\end{table}

	\section{Nuclear Refinement}
	
\begin{figure} 
		\centering\includegraphics[scale=0.52]{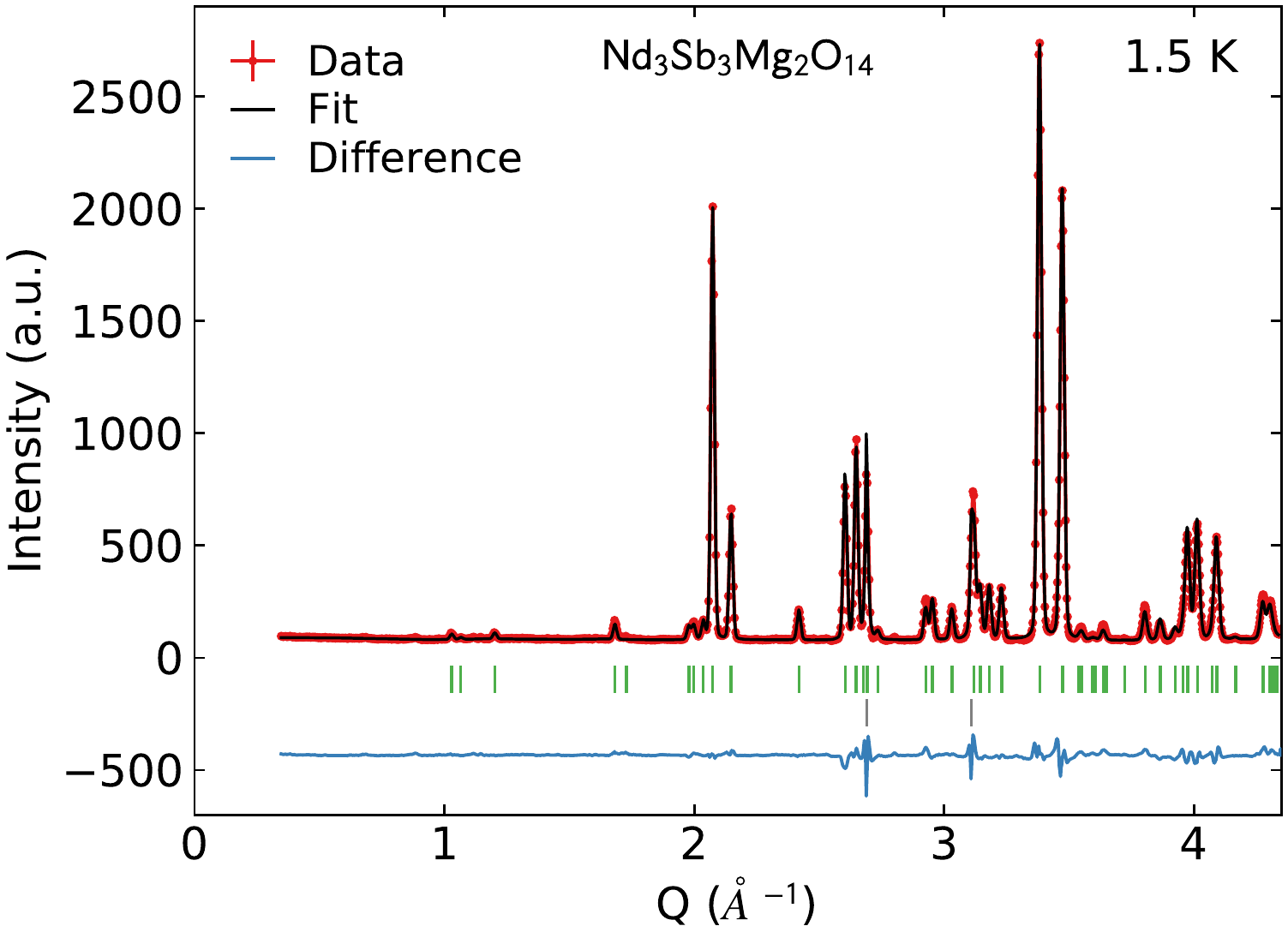}
		
		\caption{Nuclear refinement of $\rm{Nd_3Sb_3Mg_2O_{14}}$. This refinement includes peaks from the sample and from the aluminum sample can.}
		\label{flo:NuclearRefinement}
	\end{figure}

	To get accurage positions of the oxygen atoms in $\rm{Nd_3Sb_3Mg_2O_{14}}$, we performed a Rietveld refinement to neutron diffraction data. The data was taken on 10 g loose powder using the HB2A instrument at ORNL with $\lambda = 2.41$ \AA$\>$ neutrons and a 21' pre-sample monochromator. The refinement was performed using the FullProf suite \cite{Fullprof}. The data is shown in Fig. \ref{flo:NuclearRefinement} and the refined atomic positions are given in Table \ref{flo:NucRefinementStructure}. We examined Nd and Mg site mixing, but found no evidence of site mixing: all attempts to refine these gave unphysical negative mixing coefficients.
	
	\begin{table}
		\caption{Refined nuclear positions and site occupancies fractions (S.O.F.) for $\rm{Nd_3Sb_3Mg_2O_{14}}$.}
		\begin{ruledtabular}%
			\begin{tabular}{cccccc}
				atom type & label & $x$ & $y$ & $z$ & S.O.F. \tabularnewline
				\hline 
				Mg & Mg1 & 0 & 0 & 0 & 1  \tabularnewline
				Mg & Mg2 & 0 & 0 & 1/2 & 1  \tabularnewline
				Sb & Sb1 & 1/2 & 0 & 1/2 & 1  \tabularnewline
				Nd & Nd1 & 1/2 & 0 & 0 & 1  \tabularnewline
				O & O1 & 0 & 0 & 0.3856(4) & 1 \tabularnewline
				O & O2 & 0.5341(2) & 0.4660(2) & 0.1452(1) & 1 \tabularnewline
				O & O3 & 0.1441(2) & 0.8560(2) &-0.0579(2) & 1 \tabularnewline
			\end{tabular}\end{ruledtabular}
			\label{flo:NucRefinementStructure}
		\end{table}
		
		\section{Computational Methods}
		
		\subsection{PyCrystalField}
		
		The fits to the CEF Hamiltonian were performed using the PyCrystalField software package \cite{PyCrystalField}, available for download at https://github.com/asche1/PyCrystalField. This software package was written for this project.
		
		PyCrystalField contains a python library of Stevens Operators, tesseral harmonics, and physical constants for calculating the single-ion crystal Hamiltonian of a point charge model. It calculates eigenvectors and eigenvalues for a given Hamiltonian, magnetic susceptibility, directional magnetization, and the $Q$ and $\Delta E$ dependent neutron spectrum using the dipole approximation and with an arbitrary $\Delta E$ dependent resolution function. It has the capability to fit either the CEF parameters  $B_n^m$ or the effective charges of a point charge model by minimizing a user-provided global $\chi^2$ function; in this way, the user may fit any relevant data (susceptibility, neutron spectrum, magnetization, or transition energies) in any format. The minimization routines used are those in the scipy.optimize package.
		
		\subsection{Susceptibility Examination}
		
\begin{figure} 
			\centering\includegraphics[scale=0.52]{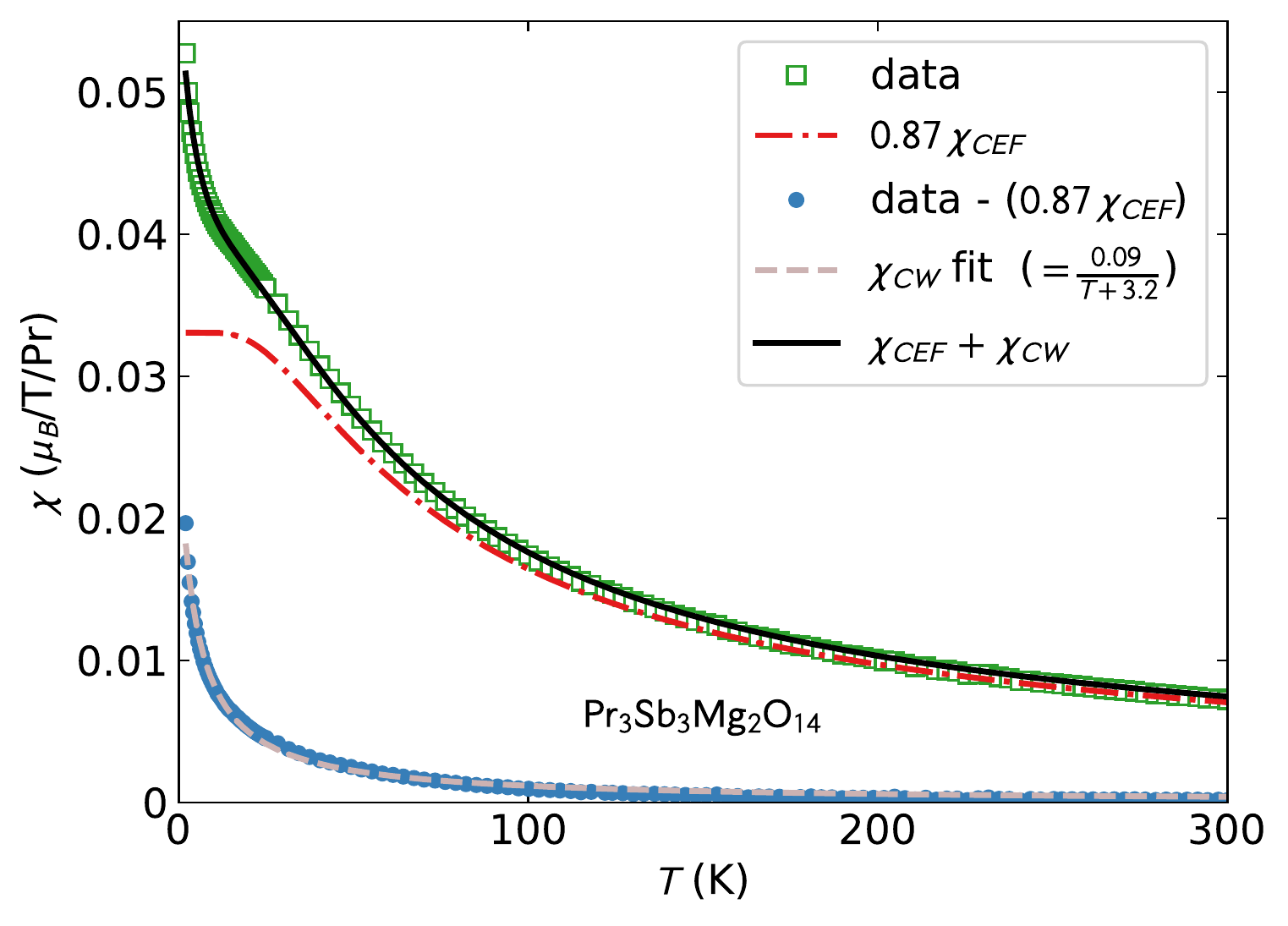}
			
			\caption{Susceptibility data and fitted impurity model for $\rm{Pr_3Sb_3Mg_2O_{14}}$. The blue data is the $\rm{Pr_3Sb_3Mg_2O_{14}}$ data minus the scaled calculated CEF susceptibility. This subtracted data, which we take to be the impurity contribution, fits almost perfectly to a Curie-Weiss law.
			}
			\label{flo:SusceptibilityImpurityFit}
		\end{figure}

		As noted in the text, the low-temperature deviation in $\rm{Pr_3Sb_3Mg_2O_{14}}$ suggests the presence of non-singlet impurities. To characterize this, we fit the susceptibility to a model where $\chi = x\> \chi_{CEF} + (1-x)\> \frac{C}{T-\theta_C}$, where $0 < x < 1$. The fit is shown in Fig. \ref{flo:SusceptibilityImpurityFit}. The fitted parameters are $x=0.871(1)$, $C=0.709(6)$, and $\theta_C = -3.2(1)$~K, indicating a 12.9\% Curie-Weiss contribution with $\mu_{eff}=1.780(7)\>\mu_B$.
		
		To be sure we had the correct explanation for the susceptibility discrepancy, we considered two alternatives: sample diamagnetism, and higher multiplet CEF levels.
		
\begin{figure} 
			\centering\includegraphics[scale=0.49]{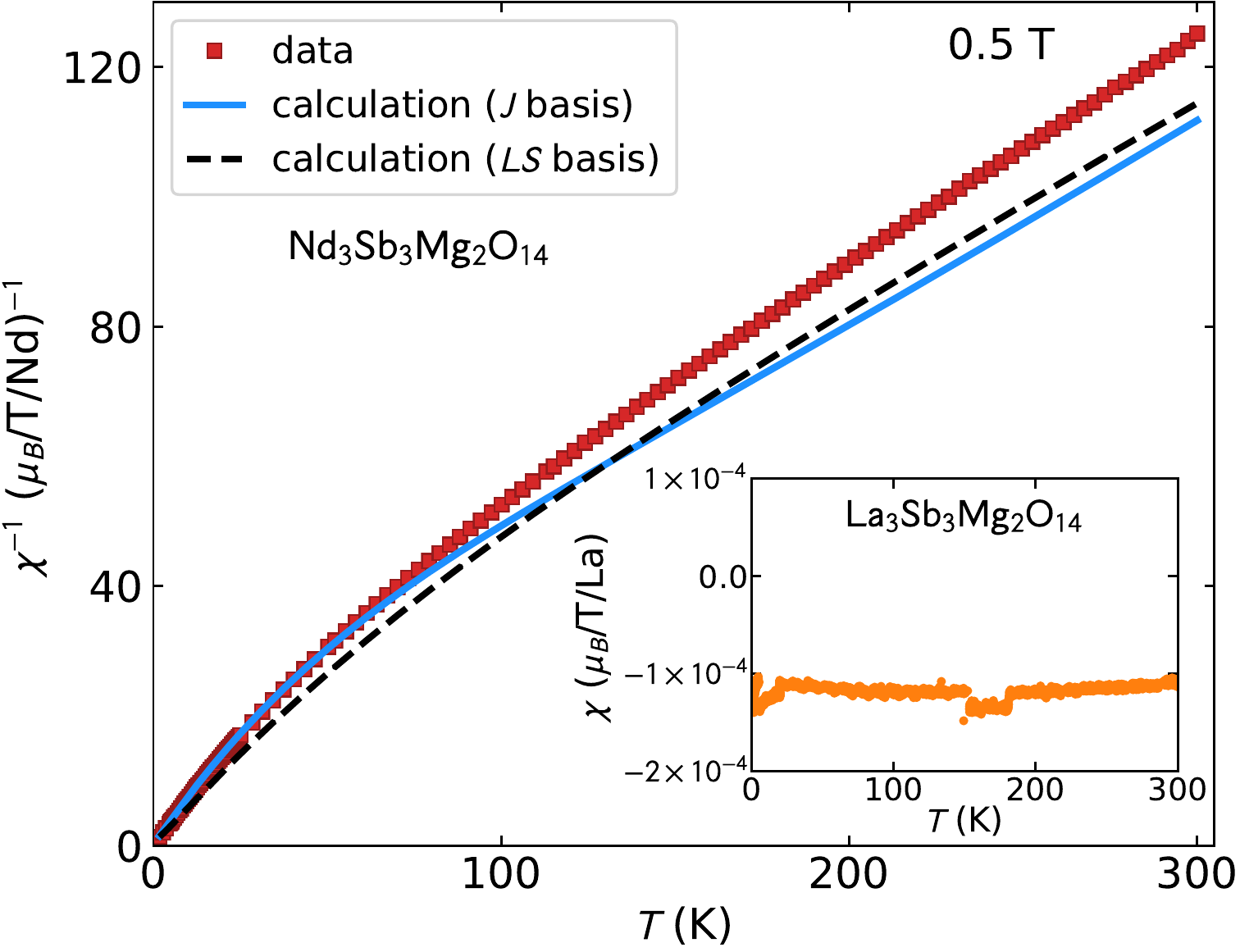}
			
			\caption{Comparison between measured susceptibility and susceptibility calculated from the intermediate coupling ($LS$ basis) and strong coupling ($J$ basis) fitted CEF Hamiltonians for $\rm{Nd_3Sb_3Mg_2O_{14}}$. Both schemes yield similar results, and neither calculation matches the data at high temperatures. The inset shows the susceptibility of nonmagnetic analogue $\rm{La_3Sb_3Mg_2O_{14}}$, giving a diamagnetic $\chi_0 = -0.0001$ ($\mu_B$/T/ion) which is too small to account for the discrepancy.
			}
			\label{flo:IntermediateSusceptibility}
		\end{figure}

		This discrepancy could be explained by a temperature-independent offset $\chi_0=-10^{-3}$ ($\mu_B$/T/ion) from sample diamagnetism. However, the diamagnetic $\chi_0$ from the non-magnetic analogue $\rm La_3 Sb_3 Mg_2 O_{14}$ (which should be similar to the $\rm Nd_3 Sb_3 Mg_2 O_{14}$ diamagnetism) is $-10^{-4}$ ($\mu_B$/T/ion)---an order of magnitude too small (see the inset in Fig. \ref{flo:IntermediateSusceptibility}). 
		
		We finally attempted to account for the susceptibility discrepancy by including higher multiplet mixing in the CEF Hamiltonian. For Nd$3+$ the first excited multiplet ($J=11/2$) is only at 230 meV \cite{Nd3_Multiplets}, so it could conceivably effect the magnetism at high temperatures. To test this we re-fit and analyzed the data using an intermediate coupling scheme: including spin-orbit coupling and calculating eigenkets in the $LS$ basis rather than the $J$ basis. Details of this calculation are given below. The results accounted for neutron data well but the calculated susceptibility, as shown in Fig. \ref{flo:IntermediateSusceptibility}, is not significantly different. We conclude that the deviation from measured susceptibility is not due to higher multiplet mixing.
		
		This leaves sample impurities as a reasonable explanation for the deviation of measured susceptibility from susceptibility calculated from crystal field levels alone.
		
		\subsection{Intermediate Coupling Scheme}
		
		Ordinarily, crystal field interactions in rare earth ions are treated as a perturbation to spin-orbit coupling, such that the CEF interacts with an effective spin $J=L+S$ in the $|J,m_J \rangle$ basis (this is called the "weak coupling scheme" \cite{AbragamBleaney}). However, when the energy scale of the next $J$ multiplet is close enough to the CEF energy levels, this approximation is no longer valid. In that case, the Hamiltonian needs to be calculated in the $|L,S,m_L,m_S\rangle$ basis (the "intermediate coupling scheme") to account for spin-orbit coupling.
		
		PyCrystalField calculates the CEF Hamiltonian in the intermediate coupling scheme by expressing the crystal fields as interacting the orbital angular momentum $L$ (CEFs, being electrostatic, are not coupled to $S$), and adding spin orbit coupling $\mathcal{H}_{SOC}=\lambda S\cdot L$ non-perturbatively to the Hamiltonian so that 
		\begin{equation}\mathcal{H} = \mathcal{H}_{SOC} + \mathcal{H}_{CEF}.\end{equation}
		From here, the eigenvalues and eigenvectors are calculated by diagonalizing the Hamiltonian. For Nd$^{3+}$, $S=1.5$ and $L=6$ so the Hamiltonian is written as a $52\times52$ matrix.
		Neutron spectrum and susceptibility are related to $J=L+S$, so in the intermediate scheme we write $|\langle \Gamma_m|\hat J_{\perp}|\Gamma_n \rangle|^2  = |\langle \Gamma_m|\hat L_{\perp} + \hat S_{\perp}|\Gamma_n \rangle|^2 $ and $M_{\alpha} = g_J \langle J_{\alpha} \rangle = \langle L_{\alpha} + g_e S_{\alpha} \rangle$.
		
		To fit the data with the intermediate coupling scheme we re-calculated the point-charge model in the $|L,S,m_L,m_S\rangle$ basis, using the method outlined in ref. \cite{Stevens1952} to calculate the $\theta_n$ in the new basis. From there, we performed an effective point-charge fit, and then a fit directly to the CEF parameters just the same as in the $J$ basis. The resulting CEF parameters are listed in Table \ref{flo:NdMg_CEF_params}. We do not list the eigenkets of the intermediate coupling fit because they are simply too long, and the calculations do not significantly differ from calculations in the $J$ basis.
		
		PyCrystalField's accuracy in the intermediate coupling regime was tested by taking the CEF parameters $B_n^m$ from the original fit in the $J$ basis, inserting them into the Hamiltonian in the $LS$ basis, and then setting the spin orbit coupling parameter $\lambda$ to a very high value (thousands of eV). In the limit where $\lambda \rightarrow \infty$, the $LS$ basis calculations should be identical to the $J$ basis calculations. This is what we observe: when $\lambda$ becomes very large, the eigenvalues, neutron spectrum, and calculated susceptibility are identical to the results from the $J$ basis. Thus, we are confident that the intermediate coupling calculations are accurate.
		
		\section{Fit Results: CEF Parameters and Eigenstates}

		The fitted CEF parameters and final eigenstates for $\rm{Nd_3Sb_3Mg_2O_{14}}$ are given in Tables \ref{flo:NdMg_CEF_params} - \ref{flo:NdMgEigenvectors}. The fitted CEF parameters and final eigenstates for $\rm{Nd_3Sb_3Zn_2O_{14}}$ are given in Tables \ref{flo:NdZn_CEF_params} - \ref{flo:NdZnEigenvectors}, with constant Q cuts for all energies and temperatures are shown in Fig. \ref{flo:NdZn_Qcuts_full}. The fitted CEF parameters and final eigenstates for $\rm{Pr_3Sb_3Mg_2O_{14}}$ are given in Tables \ref{flo:PrMg_CEF_params} - \ref{flo:PrMgEigenvectors_rescaled}, with constant Q cuts for all energies and temperatures shown in Fig. \ref{flo:PrMg_Qcuts_full}.
		
\begin{figure*} 
			\centering\includegraphics[scale=0.51]{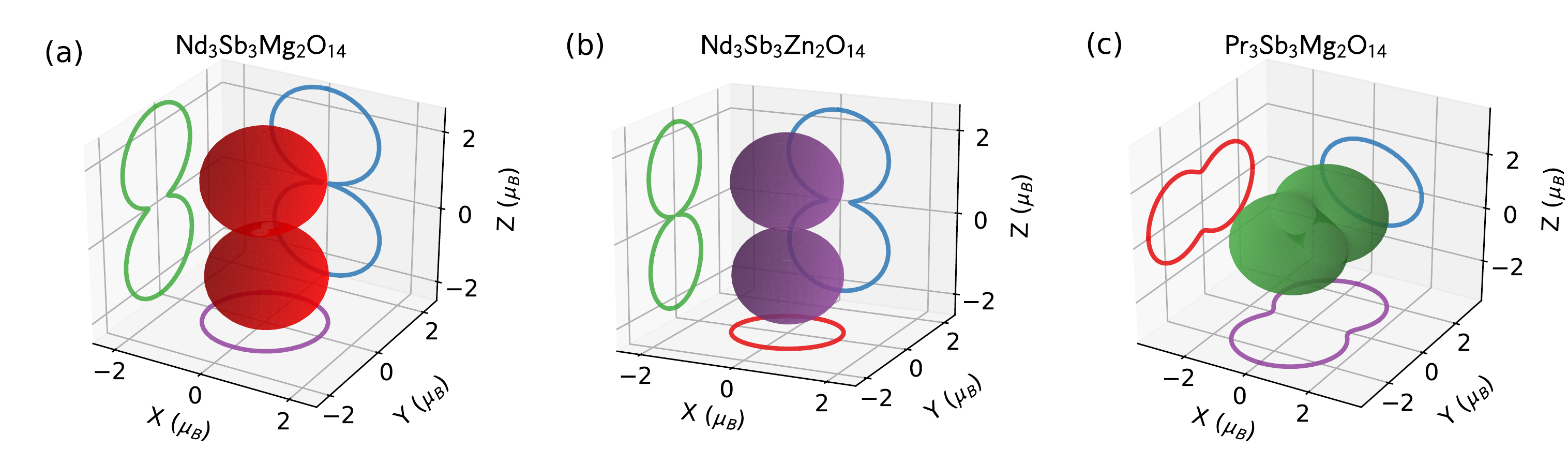}
			
			\caption{Single ion anisotropies of $\rm{Nd_3Sb_3Mg_2O_{14}}$,  $\rm{Nd_3Sb_3Zn_2O_{14}}$, and  $\rm{Pr_3Sb_3Mg_2O_{14}}$, represented by 3D plots of saturation magnetization in various directions at 2 K computed from the refined CEF parameters. The field used for $\rm{Nd_3Sb_3Mg_2O_{14}}$ and $\rm{Nd_3Sb_3Zn_2O_{14}}$ was 9 T, while the field used for $\rm{Pr_3Sb_3Mg_2O_{14}}$ was 150 T. The colored traces indicate the outline of the 3D figure along the $x$, $y$, and $z$ directions which are defined in the main text.
			}
			\label{flo:Anisotropy3D}
		\end{figure*}

\begin{figure*} 
								\centering\includegraphics[scale=0.65]{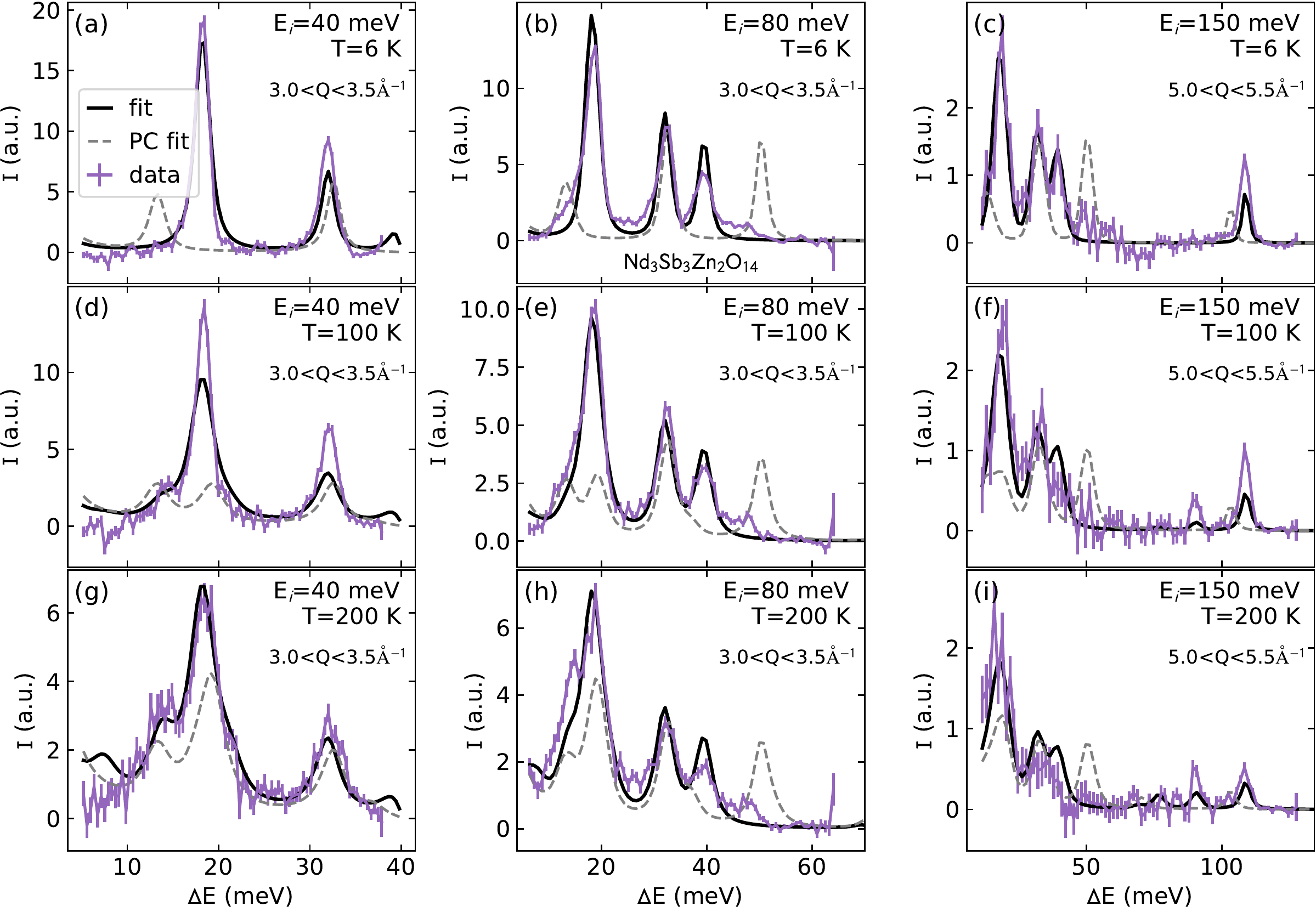}
								
								\caption{Constant Q cuts showing the results of the CEF fit to  $\rm{Nd_3Sb_3Zn_2O_{14}}$ neutron scattering data.
									The point charge fit ("PC fit") is shown with a grey dashed line, and the final fit ("fit") is shown with a solid black line.}
								\label{flo:NdZn_Qcuts_full}
							\end{figure*}

\begin{figure*} 
												\centering\includegraphics[scale=0.65]{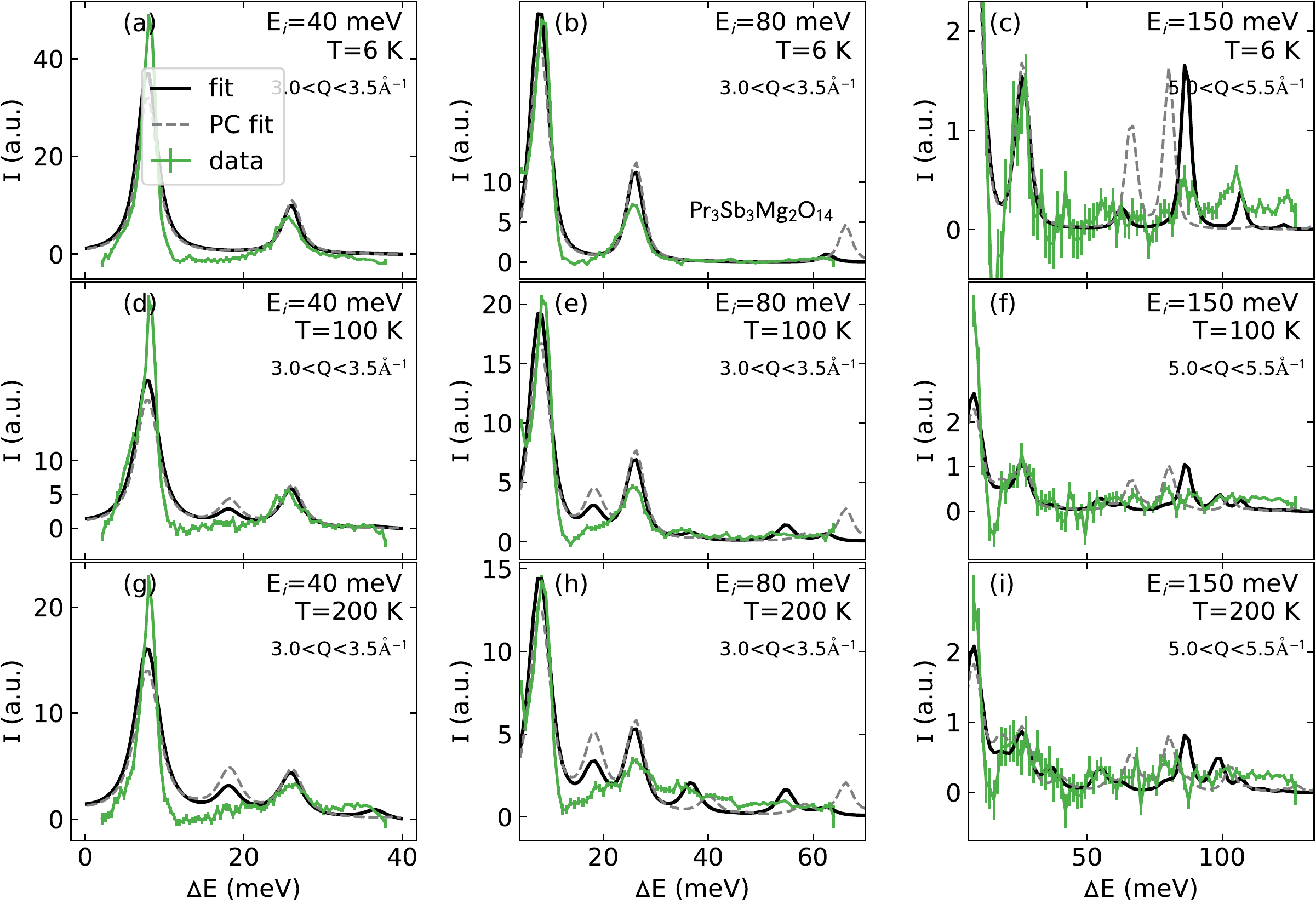}
												
												\caption{Constant Q cuts showing the results of the CEF fit to  $\rm{Pr_3Sb_3Mg_2O_{14}}$ neutron scattering data.
													The point charge fit ("PC fit") is shown with a grey dashed line, and the final fit ("fit") is shown with a solid black line.}
												\label{flo:PrMg_Qcuts_full}
											\end{figure*}

		A visual picture of magnetic anisotropy can be gained by plotting saturation $[{\bf M(B)}\cdot \hat{\bf B}] \hat{\bf B}$ in three dimensions for various field directions. These plots are shown in Fig. \ref{flo:Anisotropy3D}. These plots reveal that the anisotropy for Nd$^{3+}$ is unambiguously along $z$, while the anisotropy of Pr$^{3+}$ is along $y$. However, the field required to saturate Pr$^{3+}$ is 150 T, and the anisotropy at those high fields is highly sensitive to slight changes in the CEF Hamiltonian, so the Pr$^{3+}$ anisotropy result should be taken cautiously.
		
		\begin{table*}
			\caption{Fitted vs. Calculated CEF parameters for $\rm{Nd_3Sb_3Mg_2O_{14}}$. The first column is the result of a point charge calculation. The second column gives the result of the effective charge fit. The third column gives the result of the final fit of all CEF parameters. Columns 4-6 give the same results for the calculations in the $LS$ basis.}
			\begin{ruledtabular}
				\begin{tabular}{c|ccc|ccc}
					$B_n^m$ (meV) &Calculated PC ($J$) & PC Fit ($J$) & Final Fit ($J$)& Calculated PC ($LS$)  & PC Fit ($LS$)  & Final Fit ($LS$) \tabularnewline
					\hline 
					$ B_2^0$ & 0.24005 & -0.18422 & -0.02534 & 0.15089 & -0.09273 & -0.11967 \tabularnewline
					$ B_2^1$ & -1.59746 & -0.80356 & -1.04624 & -1.00412 & -0.60455 & -0.86278 \tabularnewline
					$ B_2^2$ & 0.14586 & -0.00991 & 0.00786 & 0.09168 & -0.11516 & 0.03186 \tabularnewline
					$ B_4^0$ & -0.03944 & -0.01916 & -0.01849 & -0.01659 & -0.00788 & -0.00593 \tabularnewline
					$ B_4^1$ & 0.00545 & 0.00299 & 0.00886 & 0.00229 & 0.0017 & 0.00243 \tabularnewline
					$ B_4^2$ & -0.01019 & -0.00441 & -0.00489 & -0.00429 & -0.00158 & -0.00168 \tabularnewline
					$ B_4^3$ & -0.33519 & -0.15385 & 0.0413 & -0.14097 & -0.06406 & -0.04794 \tabularnewline
					$ B_4^4$ & 0.02323 & 0.00993 & 0.01735 & 0.00977 & 0.00344 & 0.00525 \tabularnewline
					$ B_6^0$ & -0.00053 & -0.00027 & -0.00054 & -0.00016 & -8e-05 & -0.00014 \tabularnewline
					$ B_6^1$ & 9e-05 & 3e-05 & 1e-05 & 3e-05 & -0.0 & -0.0 \tabularnewline
					$ B_6^2$ & 0.0005 & 0.00023 & 9e-05 & 0.00015 & 7e-05 & 3e-05 \tabularnewline
					$ B_6^3$ & 0.00638 & 0.00292 & 0.00038 & 0.00188 & 0.00086 & 0.00125 \tabularnewline
					$ B_6^4$ & -0.00069 & -0.00031 & 0.00013 & -0.0002 & -9e-05 & 2e-05 \tabularnewline
					$ B_6^5$ & -0.00164 & -0.0008 & 2e-05 & -0.00048 & -0.00027 & 0.0 \tabularnewline
					$ B_6^6$ & -0.00674 & -0.00309 & -0.0028 & -0.00199 & -0.0009 & -0.00107 \tabularnewline
				\end{tabular}\end{ruledtabular}
				\label{flo:NdMg_CEF_params}
			\end{table*}

			\begin{table*}
				\caption{Eigenvectors and Eigenvalues of the calculated point-charge model for $\rm{Nd_3Sb_3Mg_2O_{14}}$.}
				\begin{ruledtabular}
					\begin{tabular}{c|cccccccccc}
						E (meV) &$| -\frac{9}{2}\rangle$ & $| -\frac{7}{2}\rangle$ & $| -\frac{5}{2}\rangle$ & $| -\frac{3}{2}\rangle$ & $| -\frac{1}{2}\rangle$ & $| \frac{1}{2}\rangle$ & $| \frac{3}{2}\rangle$ & $| \frac{5}{2}\rangle$ & $| \frac{7}{2}\rangle$ & $| \frac{9}{2}\rangle$ \tabularnewline
						\hline 
						0.000 & 0.0174 & 0.0348 & -0.0939 & 0.4554 & -0.1921 & 0.1662 & 0.1905 & -0.075 & 0.0634 & 0.8196 \tabularnewline
						0.000 & 0.8196 & -0.0634 & -0.075 & -0.1905 & 0.1662 & 0.1921 & 0.4554 & 0.0939 & 0.0348 & -0.0174 \tabularnewline
						9.990 & 0.3068 & 0.0317 & 0.2671 & 0.1325 & -0.2623 & -0.8446 & 0.038 & -0.1066 & -0.1355 & 0.0508 \tabularnewline
						9.990 & -0.0508 & -0.1355 & 0.1066 & 0.038 & 0.8446 & -0.2623 & -0.1325 & 0.2671 & -0.0317 & 0.3068 \tabularnewline
						54.600 & -0.0064 & -0.0244 & -0.2384 & 0.7924 & 0.1914 & -0.0153 & 0.2288 & -0.0135 & -0.0902 & -0.4659 \tabularnewline
						54.600 & -0.4659 & 0.0902 & -0.0135 & -0.2288 & -0.0153 & -0.1914 & 0.7924 & 0.2384 & -0.0244 & 0.0064 \tabularnewline
						94.828 & -0.115 & -0.4452 & 0.2366 & -0.0687 & 0.1997 & 0.051 & 0.2237 & -0.7968 & 0.0187 & -0.0033 \tabularnewline
						94.828 & -0.0033 & -0.0187 & -0.7968 & -0.2237 & 0.051 & -0.1997 & -0.0687 & -0.2366 & -0.4452 & 0.115 \tabularnewline
						206.713 & 0.0289 & 0.8764 & 0.1056 & -0.0245 & 0.2689 & 0.04 & 0.0407 & -0.3791 & -0.0058 & -0.0008 \tabularnewline
						206.713 & -0.0008 & 0.0058 & -0.3791 & -0.0407 & 0.04 & -0.2689 & -0.0245 & -0.1056 & 0.8764 & -0.0289 \tabularnewline
					\end{tabular}\end{ruledtabular}
					\label{flo:NdMg_InitialPC_Eigenvectors}
				\end{table*}

				\begin{table*}
					\caption{Eigenvectors and eigenvalues for effective charge PC fit of $\rm{Nd_3Sb_3Mg_2O_{14}}$, with effective charges of ($-0.999e$,  $-0.931e$,  $-0.910e$)}
					\begin{ruledtabular}
						\begin{tabular}{c|cccccccccc}
							E (meV) &$| -\frac{9}{2}\rangle$ & $| -\frac{7}{2}\rangle$ & $| -\frac{5}{2}\rangle$ & $| -\frac{3}{2}\rangle$ & $| -\frac{1}{2}\rangle$ & $| \frac{1}{2}\rangle$ & $| \frac{3}{2}\rangle$ & $| \frac{5}{2}\rangle$ & $| \frac{7}{2}\rangle$ & $| \frac{9}{2}\rangle$ \tabularnewline
							\hline 
							0.000 & -0.028 & -0.0055 & 0.0315 & -0.2853 & -0.0015 & -0.0617 & -0.1058 & -0.004 & -0.0535 & -0.9481 \tabularnewline
							0.000 & 0.9481 & -0.0535 & 0.004 & -0.1058 & 0.0617 & -0.0015 & 0.2853 & 0.0315 & 0.0055 & -0.028 \tabularnewline
							19.585 & 0.1083 & 0.1045 & 0.1767 & 0.2068 & -0.6646 & -0.5949 & -0.0911 & -0.2941 & -0.1133 & -0.0026 \tabularnewline
							19.585 & -0.0026 & 0.1133 & -0.2941 & 0.0911 & -0.5949 & 0.6646 & 0.2068 & -0.1767 & 0.1045 & -0.1083 \tabularnewline
							36.819 & 0.1039 & -0.0136 & 0.182 & -0.7156 & -0.2044 & 0.1659 & -0.5351 & -0.1049 & 0.0889 & 0.263 \tabularnewline
							36.819 & -0.263 & 0.0889 & 0.1049 & -0.5351 & -0.1659 & -0.2044 & 0.7156 & 0.182 & 0.0136 & 0.1039 \tabularnewline
							54.941 & 0.0886 & 0.4682 & -0.1719 & 0.0568 & -0.206 & -0.0649 & -0.2277 & 0.801 & 0.0048 & -0.0019 \tabularnewline
							54.941 & 0.0019 & 0.0048 & -0.801 & -0.2277 & 0.0649 & -0.206 & -0.0568 & -0.1719 & -0.4682 & 0.0886 \tabularnewline
							105.372 & -0.0265 & -0.8624 & -0.0852 & 0.0334 & -0.2894 & -0.0282 & -0.0408 & 0.3984 & -0.0483 & 0.0008 \tabularnewline
							105.372 & -0.0008 & -0.0483 & -0.3984 & -0.0408 & 0.0282 & -0.2894 & -0.0334 & -0.0852 & 0.8624 & -0.0265 \tabularnewline
						\end{tabular}\end{ruledtabular}
						\label{flo:NdMg_PC_Eigenvectors}
					\end{table*}
					
					\begin{table*}
						\caption{Final fit eigenvectors and eigenvalues for $\rm{Nd_3Sb_3Mg_2O_{14}}$}
						\begin{ruledtabular}
							\begin{tabular}{c|cccccccccc}
								E (meV) &$| -\frac{9}{2}\rangle$ & $| -\frac{7}{2}\rangle$ & $| -\frac{5}{2}\rangle$ & $| -\frac{3}{2}\rangle$ & $| -\frac{1}{2}\rangle$ & $| \frac{1}{2}\rangle$ & $| \frac{3}{2}\rangle$ & $| \frac{5}{2}\rangle$ & $| \frac{7}{2}\rangle$ & $| \frac{9}{2}\rangle$ \tabularnewline
								\hline 
								0.000 & 0.8833 & -0.0286 & -0.0348 & 0.2094 & -0.1202 & 0.0326 & 0.3962 & 0.0355 & 0.0066 & 0.011 \tabularnewline
								0.000 & 0.011 & -0.0066 & 0.0355 & -0.3962 & 0.0326 & 0.1202 & 0.2094 & 0.0348 & -0.0286 & -0.8833 \tabularnewline
								23.179 & -0.4308 & 0.0814 & -0.3726 & 0.4168 & -0.0801 & 0.0886 & 0.6666 & 0.185 & -0.0385 & -0.0322 \tabularnewline
								23.179 & 0.0322 & -0.0385 & -0.185 & 0.6666 & -0.0886 & -0.0801 & -0.4168 & -0.3726 & -0.0814 & -0.4308 \tabularnewline
								36.360 & -0.1177 & -0.0857 & 0.4117 & 0.064 & -0.8078 & 0.3728 & -0.0209 & 0.0193 & 0.1138 & 0.0 \tabularnewline
								36.360 & 0.0 & -0.1138 & 0.0193 & 0.0209 & 0.3728 & 0.8078 & 0.064 & -0.4117 & -0.0857 & 0.1177 \tabularnewline
								43.691 & 0.1342 & 0.1343 & -0.578 & -0.1059 & -0.1466 & 0.3907 & -0.4059 & 0.5206 & -0.0932 & -0.0001 \tabularnewline
								43.691 & -0.0001 & 0.0932 & 0.5206 & 0.4059 & 0.3907 & 0.1466 & -0.1059 & 0.578 & 0.1343 & -0.1342 \tabularnewline
								110.708 & 0.0352 & 0.9551 & 0.0529 & -0.0232 & -0.0349 & 0.058 & 0.0086 & -0.2162 & 0.1782 & 0.0002 \tabularnewline
								110.708 & -0.0002 & 0.1782 & 0.2162 & 0.0086 & -0.058 & -0.0349 & 0.0232 & 0.0529 & -0.9551 & 0.0352 \tabularnewline
							\end{tabular}\end{ruledtabular}
							\label{flo:NdMgEigenvectors}
						\end{table*}


						\begin{table}
							\caption{Fitted vs. Calculated CEF parameters for $\rm{Nd_3Sb_3Zn_2O_{14}}$. The first column is the result of a point charge calculation. The second column gives the result of the effective charge fit. The final column gives the result of the final fit of all CEF parameters.}
							\begin{ruledtabular}
								\begin{tabular}{c|ccc}
									$B_n^m$ (meV) &Calculated PC & PC Fit & Final Fit \tabularnewline
									\hline 
									$ B_2^0$ & 0.43977 & -0.07696 & 0.05974 \tabularnewline
									$ B_2^1$ & 0.68385 & 0.50177 & 1.48915 \tabularnewline
									$ B_2^2$ & 0.30385 & -0.05563 & -0.10943 \tabularnewline
									$ B_4^0$ & -0.0385 & -0.01895 & -0.01655 \tabularnewline
									$ B_4^1$ & 0.00109 & -0.00076 & -0.00216 \tabularnewline
									$ B_4^2$ & -0.00901 & -0.00352 & -0.00219 \tabularnewline
									$ B_4^3$ & 0.35561 & 0.16604 & 0.0169 \tabularnewline
									$ B_4^4$ & 0.02633 & 0.01041 & 0.0119 \tabularnewline
									$ B_6^0$ & -0.00051 & -0.00026 & -0.0006 \tabularnewline
									$ B_6^1$ & -0.00027 & -9e-05 & -0.00026 \tabularnewline
									$ B_6^2$ & 0.00039 & 0.00018 & 4e-05 \tabularnewline
									$ B_6^3$ & -0.00681 & -0.00318 & 0.00105 \tabularnewline
									$ B_6^4$ & -0.00061 & -0.00027 & -5e-05 \tabularnewline
									$ B_6^5$ & 0.00091 & 0.00056 & 0.00087 \tabularnewline
									$ B_6^6$ & -0.00694 & -0.00323 & -0.00268 \tabularnewline
								\end{tabular}\end{ruledtabular}
								\label{flo:NdZn_CEF_params}
							\end{table}

							\begin{table*}
								\caption{Eigenvectors and Eigenvalues of the calculated point-charge model for $\rm{Nd_3Sb_3Zn_2O_{14}}$.}
								\begin{ruledtabular}
									\begin{tabular}{c|cccccccccc}
										E (meV) &$| -\frac{9}{2}\rangle$ & $| -\frac{7}{2}\rangle$ & $| -\frac{5}{2}\rangle$ & $| -\frac{3}{2}\rangle$ & $| -\frac{1}{2}\rangle$ & $| \frac{1}{2}\rangle$ & $| \frac{3}{2}\rangle$ & $| \frac{5}{2}\rangle$ & $| \frac{7}{2}\rangle$ & $| \frac{9}{2}\rangle$ \tabularnewline
										\hline 
										0.000 & 0.4238 & -0.0572 & -0.2182 & 0.1285 & -0.4685 & -0.5819 & 0.4015 & 0.1517 & 0.1033 & 0.0 \tabularnewline
										0.000 & 0.0 & 0.1033 & -0.1517 & 0.4015 & 0.5819 & -0.4685 & -0.1285 & -0.2182 & 0.0572 & 0.4238 \tabularnewline
										5.968 & 0.6989 & 0.0582 & 0.0987 & 0.1075 & 0.3332 & 0.4378 & 0.3924 & -0.1489 & -0.0889 & 0.0 \tabularnewline
										5.968 & 0.0 & -0.0889 & 0.1489 & 0.3924 & -0.4378 & 0.3332 & -0.1075 & 0.0987 & -0.0582 & 0.6989 \tabularnewline
										54.076 & 0.5692 & 0.0555 & 0.109 & -0.2123 & -0.0362 & -0.1151 & -0.7543 & 0.1701 & 0.0591 & 0.0 \tabularnewline
										54.076 & 0.0 & 0.0591 & -0.1701 & -0.7543 & 0.1151 & -0.0362 & 0.2123 & 0.109 & -0.0555 & 0.5692 \tabularnewline
										90.926 & 0.0 & -0.1675 & 0.7057 & -0.1752 & -0.0763 & -0.1975 & 0.1036 & -0.4523 & 0.4188 & 0.0878 \tabularnewline
										90.926 & -0.0878 & 0.4188 & 0.4523 & 0.1036 & 0.1975 & -0.0763 & 0.1752 & 0.7057 & 0.1675 & 0.0 \tabularnewline
										205.317 & 0.0 & 0.0424 & 0.3835 & -0.0274 & -0.044 & -0.2789 & 0.0237 & -0.0834 & -0.8735 & -0.0129 \tabularnewline
										205.317 & 0.0129 & -0.8735 & 0.0834 & 0.0237 & 0.2789 & -0.044 & 0.0274 & 0.3835 & -0.0424 & 0.0 \tabularnewline
									\end{tabular}\end{ruledtabular}
									\label{flo:NdZn_InitialPC_Eigenvectors}
								\end{table*}
								
								\begin{table*}
									\caption{Eigenvectors and Eigenvalues for effective charge PC fit of $\rm{Nd_3Sb_3Zn_2O_{14}}$, with effective charges of ($-1.01e$, $-0.968e$, $-0.915e$)}
									\begin{ruledtabular}
										\begin{tabular}{c|cccccccccc}
											E (meV) &$| -\frac{9}{2}\rangle$ & $| -\frac{7}{2}\rangle$ & $| -\frac{5}{2}\rangle$ & $| -\frac{3}{2}\rangle$ & $| -\frac{1}{2}\rangle$ & $| \frac{1}{2}\rangle$ & $| \frac{3}{2}\rangle$ & $| \frac{5}{2}\rangle$ & $| \frac{7}{2}\rangle$ & $| \frac{9}{2}\rangle$ \tabularnewline
											\hline 
											0.000 & 0.9194 & 0.0406 & 0.0017 & 0.1574 & 0.0629 & 0.0061 & 0.3303 & -0.0329 & 0.0023 & 0.1187 \tabularnewline
											0.000 & -0.1187 & 0.0023 & 0.0329 & 0.3303 & -0.0061 & 0.0629 & -0.1574 & 0.0017 & -0.0406 & 0.9194 \tabularnewline
											13.307 & 0.0964 & -0.1277 & 0.1715 & -0.1277 & -0.7292 & 0.5438 & -0.033 & 0.3014 & -0.0963 & -0.0 \tabularnewline
											13.307 & -0.0 & 0.0963 & 0.3014 & 0.033 & 0.5438 & 0.7292 & -0.1277 & -0.1715 & -0.1277 & -0.0964 \tabularnewline
											32.647 & -0.0315 & -0.035 & 0.1295 & 0.887 & -0.1739 & -0.0389 & -0.1765 & -0.0532 & 0.0604 & -0.3533 \tabularnewline
											32.647 & 0.3533 & 0.0604 & 0.0532 & -0.1765 & 0.0389 & -0.1739 & -0.887 & 0.1295 & 0.035 & -0.0315 \tabularnewline
											50.378 & 0.0723 & -0.4499 & -0.3708 & -0.0566 & -0.1805 & 0.1206 & -0.1593 & -0.7462 & -0.1494 & 0.0021 \tabularnewline
											50.378 & -0.0021 & -0.1494 & 0.7462 & -0.1593 & -0.1206 & -0.1805 & 0.0566 & -0.3708 & 0.4499 & 0.0723 \tabularnewline
											103.234 & 0.0178 & -0.8265 & -0.0455 & 0.0449 & 0.296 & 0.0452 & 0.0118 & 0.4033 & 0.2451 & 0.0009 \tabularnewline
											103.234 & 0.0009 & -0.2451 & 0.4033 & -0.0118 & 0.0452 & -0.296 & 0.0449 & 0.0455 & -0.8265 & -0.0178 \tabularnewline
										\end{tabular}\end{ruledtabular}
										\label{flo:NdZn_PC_Eigenvectors}
									\end{table*}
									
									\begin{table*}
										\caption{Final Fit Eigenvectors and Eigenvalues for  $\rm{Nd_3Sb_3Zn_2O_{14}}$. }
										\begin{ruledtabular}
											\begin{tabular}{c|cccccccccc}
												E (meV) &$| -\frac{9}{2}\rangle$ & $| -\frac{7}{2}\rangle$ & $| -\frac{5}{2}\rangle$ & $| -\frac{3}{2}\rangle$ & $| -\frac{1}{2}\rangle$ & $| \frac{1}{2}\rangle$ & $| \frac{3}{2}\rangle$ & $| \frac{5}{2}\rangle$ & $| \frac{7}{2}\rangle$ & $| \frac{9}{2}\rangle$ \tabularnewline
												\hline 
												0.000 & 0.8077 & 0.0611 & 0.072 & 0.2893 & -0.0012 & -0.0189 & 0.4889 & -0.1198 & 0.0279 & -0.0195 \tabularnewline
												0.000 & -0.0195 & -0.0279 & -0.1198 & -0.4889 & -0.0189 & 0.0012 & 0.2893 & -0.072 & 0.0611 & -0.8077 \tabularnewline
												18.253 & 0.5333 & 0.1023 & -0.0497 & -0.4258 & -0.0601 & 0.1161 & -0.5039 & 0.4947 & -0.0725 & 0.0202 \tabularnewline
												18.253 & 0.0202 & 0.0725 & 0.4947 & 0.5039 & 0.1161 & 0.0601 & -0.4258 & 0.0497 & 0.1023 & -0.5333 \tabularnewline
												32.029 & 0.1793 & -0.1484 & -0.1206 & -0.0686 & -0.4458 & 0.1985 & -0.3953 & -0.7241 & 0.0887 & 0.0006 \tabularnewline
												32.029 & -0.0006 & 0.0887 & 0.7241 & -0.3953 & -0.1985 & -0.4458 & 0.0686 & -0.1206 & 0.1484 & 0.1793 \tabularnewline
												39.481 & -0.1589 & 0.0658 & 0.1925 & 0.096 & -0.6959 & 0.5088 & 0.2669 & 0.3307 & 0.0199 & -0.0004 \tabularnewline
												39.481 & -0.0004 & -0.0199 & 0.3307 & -0.2669 & 0.5088 & 0.6959 & 0.096 & -0.1925 & 0.0658 & 0.1589 \tabularnewline
												108.776 & -0.0708 & 0.9718 & -0.1252 & -0.0119 & 0.0038 & 0.0355 & -0.02 & -0.1798 & -0.028 & 0.0007 \tabularnewline
												108.776 & -0.0007 & -0.028 & 0.1798 & -0.02 & -0.0355 & 0.0038 & 0.0119 & -0.1252 & -0.9718 & -0.0708 \tabularnewline
											\end{tabular}\end{ruledtabular}
											\label{flo:NdZnEigenvectors}
										\end{table*}


										\begin{table}
											\caption{Fitted vs. Calculated CEF parameters for $\rm{Pr_3Sb_3Mg_2O_{14}}$. The first column is the result of a point charge calculation. The second column gives the result of the effective charge fit. The final column gives the result of the final fit of all CEF parameters.}
											\begin{ruledtabular}
												\begin{tabular}{c|ccc}
													$B_n^m$ (meV) &Calculated PC & PC Fit & Final Fit \tabularnewline
													\hline 
													$ B_2^0$ & 0.80689 & 0.33959 & 0.40799 \tabularnewline
													$ B_2^1$ & -5.71335 & -1.05144 & -4.05166 \tabularnewline
													$ B_2^2$ & 0.60658 & 1.57933 & 2.14389 \tabularnewline
													$ B_4^0$ & -0.11491 & -0.04644 & -0.05468 \tabularnewline
													$ B_4^1$ & 0.0153 & -0.00087 & 0.00182 \tabularnewline
													$ B_4^2$ & -0.03085 & -0.01614 & -0.09819 \tabularnewline
													$ B_4^3$ & -0.97192 & -0.38774 & -0.32064 \tabularnewline
													$ B_4^4$ & 0.07128 & 0.03898 & 0.11791 \tabularnewline
													$ B_6^0$ & 0.00104 & 0.00042 & -0.0004 \tabularnewline
													$ B_6^1$ & -0.00022 & -0.00022 & -0.00081 \tabularnewline
													$ B_6^2$ & -0.00101 & -0.00043 & -0.00051 \tabularnewline
													$ B_6^3$ & -0.01248 & -0.005 & 3e-05 \tabularnewline
													$ B_6^4$ & 0.00141 & 0.00062 & 0.00246 \tabularnewline
													$ B_6^5$ & 0.00324 & 0.00079 & -0.00771 \tabularnewline
													$ B_6^6$ & 0.01322 & 0.00534 & 0.00543 \tabularnewline
												\end{tabular}\end{ruledtabular}
												\label{flo:PrMg_CEF_params}
											\end{table}
											
											\begin{table*}
												\caption{Eigenvectors and Eigenvalues of the calculated point-charge model for $\rm{Pr_3Sb_3Mg_2O_{14}}$.}
												\begin{ruledtabular}
													\begin{tabular}{c|ccccccccc}
														E (meV) &$|-4\rangle$ & $|-3\rangle$ & $|-2\rangle$ & $|-1\rangle$ & $|0\rangle$ & $|1\rangle$ & $|2\rangle$ & $|3\rangle$ & $|4\rangle$ \tabularnewline
														\hline 
														0.000 & 0.0211 & 0.1351 & 0.0143 & 0.0677 & -0.9762 & -0.0677 & 0.0143 & -0.1351 & 0.0211 \tabularnewline
														27.085 & 0.481 & -0.0515 & -0.0064 & -0.5137 & -0.0648 & 0.5137 & -0.0064 & 0.0515 & 0.481 \tabularnewline
														34.102 & -0.501 & 0.0301 & -0.1321 & 0.4802 & -0.0 & 0.4802 & 0.1321 & 0.0301 & 0.501 \tabularnewline
														172.139 & -0.5001 & 0.097 & 0.1115 & -0.4758 & -0.0575 & 0.4758 & 0.1115 & -0.097 & -0.5001 \tabularnewline
														175.477 & 0.4715 & -0.1299 & 0.0341 & 0.5095 & -0.0 & 0.5095 & -0.0341 & -0.1299 & -0.4715 \tabularnewline
														258.296 & 0.0149 & 0.4974 & -0.4939 & -0.0372 & 0.1187 & 0.0372 & -0.4939 & -0.4974 & 0.0149 \tabularnewline
														267.648 & 0.1481 & 0.2102 & -0.6575 & -0.0394 & 0.0 & -0.0394 & 0.6575 & 0.2102 & -0.1481 \tabularnewline
														320.705 & -0.1336 & -0.4715 & -0.4933 & -0.0616 & -0.1593 & 0.0616 & -0.4933 & 0.4715 & -0.1336 \tabularnewline
														354.693 & -0.0683 & -0.6618 & -0.2215 & -0.0907 & 0.0 & -0.0907 & 0.2215 & -0.6618 & 0.0683 \tabularnewline
													\end{tabular}\end{ruledtabular}
													\label{flo:PrMg_InitialPC_Eigenvectors}
												\end{table*}
												
												\begin{table*}
													\caption{Eigenvectors and Eigenvalues for the effective charge PC fit of $\rm{Pr_3Sb_3Mg_2O_{14}}$, with effective charges of ($-0.805e$  $-0.736e$,  $-0.836e$)}
													\begin{ruledtabular}
														\begin{tabular}{c|ccccccccc}
															E (meV) &$|-4\rangle$ & $|-3\rangle$ & $|-2\rangle$ & $|-1\rangle$ & $|0\rangle$ & $|1\rangle$ & $|2\rangle$ & $|3\rangle$ & $|4\rangle$ \tabularnewline
															\hline 
															0.000 & -0.1362 & -0.1213 & -0.109 & 0.1183 & 0.939 & -0.1183 & -0.109 & 0.1213 & -0.1362 \tabularnewline
															7.877 & 0.3911 & -0.0156 & -0.0823 & -0.5597 & 0.2313 & 0.5597 & -0.0823 & 0.0156 & 0.3911 \tabularnewline
															26.125 & 0.5466 & 0.0405 & 0.0777 & -0.44 & 0.0 & -0.44 & -0.0777 & 0.0405 & -0.5466 \tabularnewline
															66.282 & 0.5526 & -0.1067 & -0.1292 & 0.4081 & -0.0001 & -0.4081 & -0.1292 & 0.1067 & 0.5526 \tabularnewline
															80.287 & 0.4353 & -0.1942 & -0.0029 & 0.5223 & 0.0 & 0.5223 & 0.0029 & -0.1942 & -0.4353 \tabularnewline
															110.772 & -0.0715 & -0.1053 & 0.6951 & 0.0242 & 0.0 & 0.0242 & -0.6951 & -0.1053 & 0.0715 \tabularnewline
															113.869 & -0.0114 & 0.584 & -0.3949 & 0.0431 & 0.0451 & -0.0431 & -0.3949 & -0.584 & -0.0114 \tabularnewline
															135.415 & -0.1516 & -0.3641 & -0.5557 & -0.0658 & -0.2504 & 0.0658 & -0.5557 & 0.3641 & -0.1516 \tabularnewline
															148.379 & 0.0818 & 0.6705 & 0.1036 & 0.1817 & 0.0 & 0.1817 & -0.1036 & 0.6705 & -0.0818 \tabularnewline
														\end{tabular}\end{ruledtabular}
														\label{flo:PrMg_PC_Eigenvectors}
													\end{table*}
													
													\begin{table*}
														\caption{Final Eigenvectors and Eigenvalues for $\rm{Pr_3Sb_3Mg_2O_{14}}$. As required for singlets, $\langle j_x \rangle = \langle j_y \rangle = \langle j_z \rangle =0$ for all states.}
														\begin{ruledtabular}
															\begin{tabular}{c|ccccccccc}
																E (meV) &$|-4\rangle$ & $|-3\rangle$ & $|-2\rangle$ & $|-1\rangle$ & $|0\rangle$ & $|1\rangle$ & $|2\rangle$ & $|3\rangle$ & $|4\rangle$ \tabularnewline
																\hline 
																0.000 & 0.3301 & 0.0176 & 0.2111 & -0.0663 & -0.8268 & 0.0663 & 0.2111 & -0.0176 & 0.3301 \tabularnewline
																7.857 & 0.1378 & -0.0761 & 0.0021 & -0.6723 & 0.2156 & 0.6723 & 0.0021 & 0.0761 & 0.1378 \tabularnewline
																25.982 & 0.5732 & -0.0511 & 0.294 & -0.287 & 0.0 & -0.287 & -0.294 & -0.0511 & -0.5732 \tabularnewline
																62.734 & 0.5925 & -0.1582 & -0.1293 & 0.198 & 0.3686 & -0.198 & -0.1293 & 0.1582 & 0.5925 \tabularnewline
																86.585 & -0.3312 & 0.2206 & 0.1273 & -0.5704 & 0.0 & -0.5704 & -0.1273 & 0.2206 & 0.3312 \tabularnewline
																106.952 & 0.2236 & 0.1408 & -0.6201 & -0.2138 & -0.0 & -0.2138 & 0.6201 & 0.1408 & -0.2236 \tabularnewline
																122.734 & -0.0693 & -0.3987 & 0.5582 & 0.0651 & 0.2022 & -0.0651 & 0.5582 & 0.3987 & -0.0693 \tabularnewline
																181.173 & 0.1272 & 0.5567 & 0.3566 & 0.0131 & 0.3052 & -0.0131 & 0.3566 & -0.5567 & 0.1272 \tabularnewline
																197.149 & -0.1083 & -0.6549 & -0.1133 & -0.2157 & -0.0 & -0.2157 & 0.1133 & -0.6549 & 0.1083 \tabularnewline
															\end{tabular}\end{ruledtabular}
															\label{flo:PrMgEigenvectors}
														\end{table*}
														
														\begin{table*}
															\caption{Eigenvectors and Eigenvalues for $\rm{Pr_3Sb_3Mg_2O_{14}}$ obtained by rescaling the Nd$^{3+}$ CEF parameters.}
															\begin{ruledtabular}
																\begin{tabular}{c|ccccccccc}
																	E (meV) &$|-4\rangle$ & $|-3\rangle$ & $|-2\rangle$ & $|-1\rangle$ & $|0\rangle$ & $|1\rangle$ & $|2\rangle$ & $|3\rangle$ & $|4\rangle$ \tabularnewline
																	\hline 
																	0.000 & 0.0814 & 0.0325 & -0.1446 & -0.0809 & 0.9642 & 0.0809 & -0.1446 & -0.0325 & 0.0814 \tabularnewline
																	32.860 & 0.0109 & -0.0451 & -0.0799 & 0.6979 & 0.0943 & -0.6979 & -0.0799 & 0.0451 & 0.0109 \tabularnewline
																	105.320 & -0.0703 & 0.04 & 0.0785 & -0.6981 & 0.0 & -0.6981 & -0.0785 & 0.04 & 0.0703 \tabularnewline
																	124.743 & 0.648 & -0.2289 & 0.155 & -0.061 & 0.0 & -0.061 & -0.155 & -0.2289 & -0.648 \tabularnewline
																	126.023 & -0.643 & 0.2834 & -0.0559 & 0.0168 & 0.0756 & -0.0168 & -0.0559 & -0.2834 & -0.643 \tabularnewline
																	163.849 & -0.2515 & -0.3555 & 0.5531 & 0.0671 & -0.0 & 0.0671 & -0.5531 & -0.3555 & 0.2515 \tabularnewline
																	164.076 & 0.2598 & 0.5458 & -0.3433 & 0.0043 & -0.1829 & -0.0043 & -0.3433 & -0.5458 & 0.2598 \tabularnewline
																	189.555 & -0.1091 & -0.5654 & -0.4049 & -0.067 & -0.0 & -0.067 & 0.4049 & -0.5654 & 0.1091 \tabularnewline
																	189.810 & 0.1111 & 0.3444 & 0.5931 & 0.0784 & 0.149 & -0.0784 & 0.5931 & -0.3444 & 0.1111 \tabularnewline
																\end{tabular}\end{ruledtabular}
																\label{flo:PrMgEigenvectors_rescaled}
															\end{table*}

\end{document}